\documentclass[nofootinbib,reprint,showkeys,showpacs,amsmath,amssymb,aps]{revtex4-1}
\usepackage{graphicx}
\usepackage[utf8,latin1]{inputenc}
\usepackage{dcolumn}
\usepackage{xcolor}
\usepackage{bm}
\usepackage{hyperref}
\usepackage{amsmath}
\newcommand{\qm}[1]{``#1''}
\usepackage{hyperref}
\usepackage{colortbl}
\hypersetup{colorlinks, linkcolor={red},citecolor={blue},urlcolor={blue}}  
\newcommand\ChangeRT[1]{\noalign{\hrule height #1}}
\def\apjl{ApJL }

\def\apj{ApJ }

\def\aap{A\&A }

\def\mnras{MNRAS }

\def\ssr{SSR }
\def\prd{PRD }

\begin{document}

\title[Testing wormhole solutions in extended gravity through the Poynting-Robertson effect]{Testing wormhole solutions in extended gravity through the Poynting-Robertson effect}

\author{Vittorio De Falco$^{1}$}\email{vittorio.defalco@physics.cz}
\author{Emmanuele Battista$^{2,3}$}\email{ emmanuele.battista@kit.edu}
\author{Salvatore Capozziello$^{4,5,6}$}\email{capozziello@unina.it}
\author{Mariafelicia De Laurentis$^{4,5,7}$\vspace{0.5cm}}\email{mariafelicia.delaurentis@unina.it}

\affiliation{$^1$Department of Mathematics and Applications \qm{R. Caccioppoli}, University of Naples Federico II, Via Cintia, 80126 Naples, Italy,\\
$2$ Institute for Theoretical Physics, Karlsruhe Institute of Technology (KIT), 76128 Karlsruhe, Germany\\
$3$ Institute for Nuclear Physics, Karlsruhe Institute of Technology (KIT), Hermann-von-Helmholtz-Platz 1, 76344 Eggenstein-Leopoldshafen, Germany\\
$^4$ Universit\`{a} degli studi di Napoli \qm{Federico II}, Dipartimento di Fisica \qm{Ettore Pancini}, Complesso Universitario di Monte S. Angelo, Via Cintia Edificio 6, 80126 Napoli, Italy\\
$^5$ Istituto Nazionale di Fisica Nucleare, Sezione di Napoli, Complesso Universitario di Monte S. Angelo, Via Cintia Edificio 6, 80126 Napoli, Italy\\
$^6$Scuola Superiore Meridionale, Universit\`{a}  di Napoli \qm{Federico II},  Largo San Marcellino 10, 80138 Napoli, Italy\\
$^7$ Lab.Theor.Cosmology,Tomsk State University of Control Systems and Radioelectronics(TUSUR), 634050 Tomsk, Russia}

\date{\today}

\begin{abstract}
We develop a model-independent procedure to single out static and spherically symmetric wormhole solutions based on the general relativistic Poynting-Robertson effect and the extension of the ray-tracing formalism in generic static and spherically symmetric wormhole metrics. Simulating the flux emitted by the Poynting-Robertson critical hypersurface (i.e., a stable structure where gravitational and radiation forces attain equilibrium) or also from another X-ray source in these general geometrical environments toward a distant observer, we are able to reconstruct, only locally to the emission region, the wormhole solutions which are in agreement with the high-energy astrophysical observational data. This machinery works only if wormhole evidences have been detected. Indeed, in our previous paper we showed how the Poynting-Robertson critical hypersurfaces can be located in regions of strong gravitational field and become valuable astrophysical probe to observationally search for wormholes' existence. As examples, we apply our method to selected wormhole solutions in different extended theories of gravity by producing lightcurves, spectra, and images of an accretion disk. In addition, the present approach may constitute a procedure to also test the theories of gravity. Finally, we discuss the obtained results and draw the conclusions.
\end{abstract}

\maketitle

\section{Introduction}
\label{sec:intro}
The recent and future great amount of high-energy astrophysical observational data (e.g., International Gamma-Ray Astrophysics Laboratory (INTEGRAL) \cite{Winkler2003}, Swift \cite{Burrows2005}, XMM-Newton \cite{Beckwith2004,Tomsick2014}, Event Horizon Telescope (EHT) \cite{Chael2016,Akiyama:2019cqa}, Advanced Telescope for High-ENergy Astrophysics (ATHENA) \cite{Barcons2017}, enhanced X-ray Timing and Polarimetry mission (eXTP) \cite{Zhang2016}, Imaging x-ray Polarimetry Explorer (IXPE) \cite{Soffitta2013}, Laser Interferometer Gravitational-Wave Observatory (LIGO) and Virgo \cite{Gourgoulhon2019,Abuter2020}) is triggering the attention of all scientific community in developing more and more advanced models and in proposing new strategies to test gravity in strong field regimes. A particular attention is reserved to  wormholes (WHs), which are exotic astrophysical objects endowed with no horizons and singularities, and a characteristic traversable bridge (WH neck) connecting two different universes or extremely far regions of spacetime \cite{Visser1995}. 

The study of these objects permits to have more insight into the fundamental physics and entails important implications, like: understanding how gravity shapes spacetime topology, learning more about the exotic matter and how quantum mechanics effects couple with gravity in strong field regimes, having natural laboratories for testing quantum gravity models and the intriguing possibility for undertaking interstellar travels. 

Both in general relativity (GR) and in extended/alternative theories of gravity \cite{Capozziello:2011et} there exists a restricted class of WH solutions which have the peculiar propriety to be \emph{black hole (BH) mimickers}, meaning that they possess geometrical proprieties, which make them observationally similar to BHs \cite{Damour2007,Bambi2012,Bambi2013}. This can be one of the reasons why no observational evidence of their existence has been found so far. To this end, there are international efforts in providing feasible observational procedures to detect their signature, like: analysing the quasinormal-mode spectrum in the gravitational wave emission \cite{Cardoso2016,Konoplya2016}; producing numerical images of a thin accretion disk forming around WHs \cite{Paul2019}; the influence of gravitational fluxes, propagating from one universe to the other one through the WH neck, and influencing the accelerations of stellar objects \cite{Dai2019}; the presence of strong tidal effects close to the WH neck, strongly depending on their geometries \cite{Banerjee2019}; the absence of chaotic motions near strong gravitational field regions due to the lack of an horizon \cite{Hashimoto2017,Dalui2019}. Recently, we have proposed another strategy for the detection of WHs' existence through the general relativistic Poynting-Robertson (PR) effect. This method is based on the analysis of possible metric changes in the vicinity of a BH event horizon through the fit of observational data using the flux emitted by the PR critical hypersurface, located in such strong field regime regions, toward a distant observer \cite{Defalco2020WH}. 

The general relativistic PR effect is an important phenomenon involving the motion of relatively small test particles (e.g., dust grains, accreting matter elements, meteors, comets) around compact objects (like neutron stars or BHs), which are affected not only by the gravitational pull but also by a radiation field from an emitting source (e.g., a boundary layer around an neutron star, a hot corona around a BH, or a type-I X-ray burst on the neutron star surface). Besides having the radiation pressure opposite in direction to the gravitational attraction, there is also the radiation drag force, arising during the process of absorption and re-emission of radiation from the test particle \cite{Ballantyne2004,Ballantyne2005,Worpel2013,Ji2014,Keek2014,Worpel2015}. The latter force, known in the literature as PR effect \cite{Poynting1903,Robertson1937}, removes very efficiently energy and angular momentum from the test particle, being therefore a full-fledged dissipative effect in GR.

The PR effect in GR is steadily acquiring a prominent role in high-energy astrophysics, as it can be seen from the increasing numbers of theoretical and applicative works on such topic. Among the recent works, it is worth citing: its description in two-dimensional (2D) \cite{Bini2009,Bini2011} and three-dimensional (3D) spaces \cite{Defalco20183d,Bakala2019,Wielgus2019,Defalco2020III}, its treatment under a Lagrangian point of view \cite{DeFalco2018,Defalco2019,DeFalco2019VE,Defalco2020NA}, the proof that the critical hypersurfaces are stable configurations, different models of accretion disk dynamics affected by intense thermonuclear explosions \cite{Walker1989,Walker1992,Lancova2017,Fragile2020}. 

In this work, we want to complement our previous analysis \cite{Defalco2020WH}, where we developed an astrophysical strategy to search for the WH existence. Indeed, we proved that if metric changes occur in the transition surface layer (located between the BH event horizon and the photon sphere), then a WH could exist. In light of this hypothesis, we aim at distinguishing among the various WH solutions the most suitable for fitting the observational data. To this end, we extend the ray-tracing treatment to generic Morris-Thorne-like metrics, which represent the new geometrical framework of our current analysis.  On the other hand, since the WH metric parameters strictly depend on the underlying  gravity  theory (see, e.g. \cite{Visser1989,Barcelo1999,Bohmer2012,Anchordoqui:2000ut,Bahamonde:2016jqq,Capozziello2020}), this approach could be also considered as a further method to test extended/alternative gravity models.

The paper is organized as follows: in Sec. \ref{sec:geometry} we introduce  the Morris-Throne-like metric and we derive  the ray-tracing equations in such generic WH spacetimes; in Sec. \ref{sec:PRE} the general relativistic PR effect and its critical hypersurface equations are recalled; in Sec. \ref{sec:flux} we develop the equations to calculate the flux from an emitting source toward a distant observer; in Sec. \ref{sec:examples}, applications to some WH solutions in GR and extended/alternative theories of gravity are considered; in Sec. \ref{sec:end} we discuss our results and draw conclusions.

\section{Ray-tracing in spherically symmetric and static wormhole spacetimes}
\label{sec:geometry}

\subsection{Morris-Thorne metric structure}
\label{sec:metric}
\begin{figure*}
\centering
\includegraphics[scale=0.47]{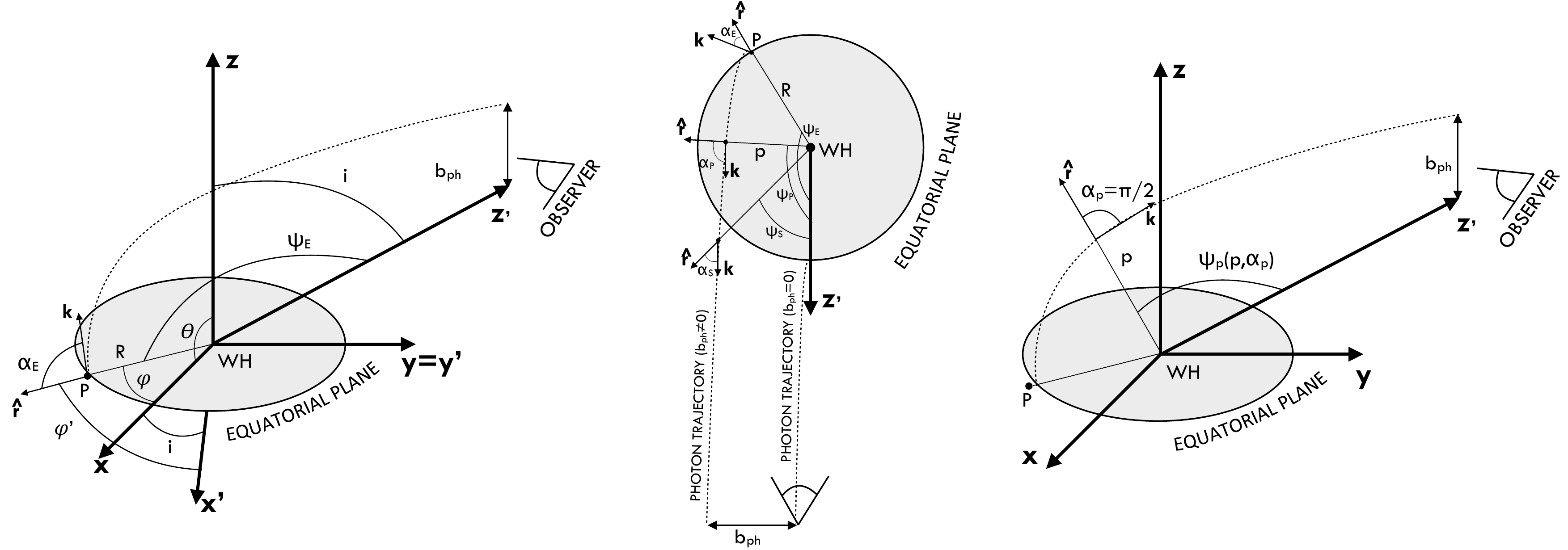}
\caption{Left panel: Ray-tracing geometry. Middle panel: Symmetrization process. Right panel: periastron geometry.}
\label{fig:Fig1}
\end{figure*}

A static and spherically symmetric WH can be generally described by a Morrison-Thorne-like metric \cite{Morris1988}, which in spherical coordinates $\left\{t,r,\theta,\psi\right\}$ and geometrical units $G=c=1$ reads as (see Fig. \ref{fig:Fig1})
\begin{equation} \label{eq:MTmetric}
\begin{aligned}
&ds^2=g_{\alpha\beta}dx^\alpha dx^\beta=\\
&=-e^{2\Phi(r)}dt^2+\frac{dr^2}{1-b(r)/r}+r^2(d\theta^2+\sin^2\theta d\psi^2),
\end{aligned}
\end{equation}
where $\Phi(r)$ and $b(r)$ denote  the redshift and shape functions, respectively. \emph{Equation (\ref{eq:MTmetric}), representing a family of metrics depending on the  functions $\Phi(r),b(r)$, is valid both in GR and in extended/alternative theories of gravity.} This class of metrics may also depend on other parameters, which can be closely determined with the help of specific constraints in different extended theories of gravity (see Sec. \ref{sec:examples}, for further details). In the WH case we use the Arnowitt, Deser, Misner (ADM) mass $M$, which represents the total mass of the system contained in the whole spacetime (see Ref. \cite{Visser1995}, for more details).

\subsubsection{Geometrical proprieties}
\label{sec:geo_pro_WH}
We briefly recall the relevant geometrical proprieties of the Morris-Thorne-like metrics \cite{Morris1988}. The absence of horizons and singularities requires  $\Phi(r)$ and $b(r)$ to be  smooth functions in $\mathbb{R}$ and $\Phi(r)$ to be everywhere finite; furthermore, the condition $1-b(r)/r\ge0$ allows to define a finite proper radial distance; the flaring outward condition requires that  $b^{\prime}(r) < b(r)/r$ near and at the throat, defined as the minimum radius such that $r_{\rm min}=b_0$ and $b(r_{\rm min})=b_0$; the asymptotic flatness entails $b(r)/r\to0$ and $\Phi(r)\to0$ for $r\to + \infty$. Traversability of WHs is subjected to the fulfilment of the flaring outward condition \cite{Hochberg19981,Hochberg19982,caplobo1,caplobo2} which, depending on the underlying theory of gravity, can be achieved by considering  exotic matter \cite{Hochberg1997,Bronnikov2013,Garattini2019} or topological defects \cite{Lobo2009,Harko2013,Digrezia2017}.

\subsubsection{Conserved quantities and geodesic structure}
\label{sec:CL}
The metric (\ref{eq:MTmetric}) depends  on the radial coordinate $r$ only, therefore there exist the timelike Killing vector $\boldsymbol{\partial_t}$ and (for closed orbits only) the spacelike Killing vector $\boldsymbol{\partial_\psi}$; the associated conserved quantities along all types of trajectories are, respectively, the energy $E$ and the angular momentum $L_d$ with respect to any direction $\boldsymbol{d}$ \cite{Misner1973}.

The spherical symmetry of the background geometry guarantees that the geodesic motion of particles and photons lies in one single plane termed \emph{invariant plane} \cite{Misner1973}. Therefore, we can investigate the dynamics of a photon in the invariant plane $\theta=\pi/2$, i.e.,  the equatorial plane in the coordinate system $\left\{t,r,\theta,\psi\right\}$ (see Fig. \ref{fig:Fig1}). 

\subsection{Ray-tracing equations}
\label{sec:RTE}
We derive the ray-tracing equations to describe the effects related to photon geodesics in metric (\ref{eq:MTmetric}), which are: \emph{light bending, travel time delay, and solid angle} \cite{Misner1973}. We consider the \emph{WH reference frame} $\left\{\boldsymbol{x},\boldsymbol{y},\boldsymbol{z}\right\}$, which is centred at the origin of the WH location and having the $\boldsymbol{x}$- and $\boldsymbol{y}$-axes lying in the equatorial plane, and the $\boldsymbol{z}$-axis orthogonal to the equatorial plane, see Fig. \ref{fig:Fig1}. To describe the photon emission, it is useful to employ spherical coordinates, where the radius $r$ joins the centre of the coordinates with any point in the space, the azimuthal angle $\varphi$ measured clockwise from the $\boldsymbol{x}$-axis in the equatorial plane, and the latitudinal angle $\theta$ measured from the $\boldsymbol{x}$-axis. We consider also a static and not rotating observer located at infinity, who is inclined by an angle $i$ with respect to the $\boldsymbol{z}$-axis, and the $\boldsymbol{z'}$-axis points in the observer's direction. The azimuthal angle $\psi$ is measured from the emission point to the $\boldsymbol{z'}$-axis, and it is known in the literature as \emph{light bending angle} \cite{Misner1973,Beloborodov2002,Defalco2016}.

The photon four velocity is given by $k^\alpha=dx^\alpha/ds$, where $s$ is the affine parameter along the photon trajectory, and $s\to r$ as $r\to\infty$ \cite{Beloborodov2002}. From the conservation laws (see Sec. \ref{sec:CL}), we have
\begin{equation} \label{eq:phcomps}
k_t=-E,\quad k_\theta=0,\quad k_\psi=L_d\footnote{The angular momentum $L_d$ is conserved in the plane $\boldsymbol{x}-\boldsymbol{y}$ along the direction $\boldsymbol{d}$ orthogonal to the invariant plane, where the photon trajectory lies. Similar argument is valid also for the angular momentum $L_{d,p}$ related to timelike trajectories.},
\end{equation}
The photon impact parameter $b_{\rm ph}\equiv L_d/E$ is naturally conserved along the photon trajectory. For determining $k^r$, we exploit the other conserved quantity, given by the module of the null geodesics four-velocity, namely $k^\alpha k_\alpha=0$. Without loss of generality we can set $E=1$, because one could consider the photon four-momentum for unity energy (i.e., $\tilde{k}^\alpha=k^\alpha/E$). Therefore, we obtain 
\begin{equation} \label{eq:phcompr}
\begin{aligned}
k^r&=\sqrt{\frac{-g^{tt}k_t^2-k_\psi^2g^{\psi\psi}}{g_{rr}}}\\
&=\sqrt{\left[1-\frac{b(r)}{r}\right]\left[e^{-2\Phi(r)}-\frac{b_{\rm ph}^2}{r^2}\right]}.
\end{aligned}
\end{equation}

\subsubsection{Impact parameter}
\label{sec:IP}
Equations (\ref{eq:phcomps}) and (\ref{eq:phcompr}) permit to explicitly calculate the impact parameter $b_{\rm ph}$ in terms of the emission angle $\alpha_E$, which is formed between the emitted photon four-velocity and the local radial direction (see Fig. \ref{fig:Fig1}). Using the conditions $\tan\alpha_E=\sqrt{(k^\psi k_\psi)/(k^r k_r)}$ and $\sin\alpha_E=\tan\alpha_E/\sqrt{1+\tan^2\alpha_E}$ \cite{Misner1973,Beloborodov2002}, we obtain
\begin{equation} \label{eq:bim}
b_{\rm ph}=\frac{R\sin\alpha_E}{e^{\Phi(R)}},
\end{equation}
where $R$ is the radius at which the photon is emitted. 

\subsubsection{Light bending}
\label{sec:LB}
We describe how the photon  trajectory is deflected when it is emitted in the WH spacetime. This is described by the parameter $\psi$, see Fig. \ref{fig:Fig1}. We set $\psi=0$ at the emission point $R$, and a generic angle $\psi$ at the observer location at infinity. The light bending angle $\psi$ is described by the following elliptic integral \cite{Misner1973,Beloborodov2002}
\begin{equation} \label{eq:LB}
\begin{aligned}
\psi&=\int_0^\psi d\Psi=\int_R^\infty\frac{d\Psi}{dr}dr=\int_R^\infty\frac{k^\psi}{k^r}dr=\\
&=\int_R^\infty\frac{b_{\rm ph}}{r^2}\left\{\left[1-\frac{b(r)}{r}\right]\left[e^{-2\Phi(r)}-\frac{b_{\rm ph}^2}{r^2}\right]\right\}^{-1/2}dr.
\end{aligned}
\end{equation}
This equation is valid for $0\le\alpha_E\le\pi/2$ \cite{Beloborodov2002,Defalco2016}. For every $\alpha_E$ we can determine the corresponding $\psi$ by solving the integral (\ref{eq:LB}), and viceversa for every $\psi$ we can determine the corresponding $\alpha_E$ by inverting Eq. (\ref{eq:LB}) through a numerical interpolation method \cite{Press2002}. 

\subsubsection{Turning points and symmetrization processes}
The photon trajectories reach the observer at infinity only if their impact parameters $b_{\rm ph}$ are greater than the critical impact parameter $b_c$, i.e., $b_{\rm ph}\ge b_c$. However, photons endowed with an impact parameter $b_{\rm ph}< b_c$ fall inside the WH throat, and end up in the opposite universe of their origins. In this work, we do not analyse the possible destiny of a photon crossing the WH throat, and we assume that it is not visible to the observer at infinity. In our numerical simulations we would like to give more emphasis to the WH geometrical background. Nevertheless, if an observer at infinity detected something coming from inside the WH throat, this would be a clear and distinctive signature of a WH existence, and can be better worked it out through the gathered observational data. 

We consider the null geodesic equation
\begin{equation} \label{eq:ng}
g_{\alpha\beta}\frac{dx^\alpha}{ds}\frac{dx^\beta}{ds}=0,\ \Rightarrow\ \frac{E^2}{e^{2\Phi(r)}}-\frac{\dot{r}^2}{1-\frac{b(r)}{r}}-\frac{L_d^2}{r^2}=0.
\end{equation}
The effective potential $V(r)$ is given by \cite{Chandrasekhar1983}
\begin{equation}
E^2\equiv V(r)=\frac{L_d^2 }{r^2} e^{2\Phi(r)}+\left[\frac{e^{2\Phi(r)}}{1-b(r)/r}\right]\dot{r}^2.
\end{equation}

The photon innermost stable circular orbit (ISCO), or also known as \emph{photon sphere} $r_{ps}$, is determined by imposing both that $\dot{r}=0$ (circular orbit) \cite{Misner1973,Chandrasekhar1983},
\begin{equation}
V(r)=\frac{L_d^2}{r^2}e^{2\Phi(r)},
\end{equation}
and that $[dV(r)/dr]_{r=r_{ps}}=0$ (stable and innermost) 
\begin{eqnarray}\label{photonsphere1}
r_{ps} \Phi'(r_{ps})-1=0,
\end{eqnarray}
where $\Phi'(r)=\Phi'(r)$. Therefore,
the critical impact parameter $b_c$ is determined by substituting the value of $r_{ps}$ in Eq. (\ref{eq:bim}) and imposing $\alpha_E=\pi/2$,
\begin{equation} \label{eq:crim}
b_c=\frac{r_{ps}}{e^{\Phi(r_{ps})}}.
\end{equation}
To calculate the maximum emission angle $\alpha_{\rm max}$ for which the photons do not fall inside the WH throat, we substitute Eq. (\ref{eq:crim}) in Eq. (\ref{eq:bim}), obtaining thus \cite{Defalco2016}
\begin{equation} \label{eq:AMAX1}
\alpha_{\rm max}=\pi-\arcsin\left[b_c\frac{e^{\Phi(R)}}{R}\right],
\end{equation}
which is valid for $R>r_{ps}$. Instead for emission radii $R\le r_{ps}$, the maximum emission angle is \cite{Defalco2020WH}
\begin{equation} \label{eq:AMAX2}
\alpha_{\rm max}=\arcsin\left[b_c\frac{e^{\Phi(R)}}{R}\right].
\end{equation}
We note that in the BH case, the gravitational pull is so strong under the photon sphere that the emission of light is restricted to the so-called \emph{cone of avoidance}, which shrinks more and more toward the event horizon \cite{Chandrasekhar1983,Defalco2020WH}. In the WH case, it is not known a-priori where and how the cone of avoidance extends through the WH neck up to the WH throat, because this behaviour strongly depends on the explicit expression of the function $\Phi(r)$. However, if $\Phi(r)$ is a \emph{monotone increasing function}, as it occurs in the BH case, we can apply Eq. (\ref{eq:AMAX2}) all the way down to the WH throat $b_0$, and the cone shrinks more and more, until it reaches its maximum at $b_0$.

We distinguish between \emph{direct photons}, where $0\le\alpha_E\le\pi/2$, and \emph{photons endowed with one turning point}, where $\pi/2\le\alpha_E\le\alpha_{\rm max}$ \cite{Defalco2016,Defalco2020WH}. For photons showing a turning point, the bending angle $\psi_E$, associated to the emission angle $\alpha_E$, cannot be straightforwardly calculated through Eq. (\ref{eq:LB}), because it is valid only for $0\le\alpha_E\le\pi/2$. However, by employing a \emph{symmetrization process} with respect to the emission angle $\alpha_p=\pi/2$, we can determine $0\le\alpha_S\le\pi/2$, given by $\alpha_S=\pi-\alpha_E$.
The angle $\alpha_p$ is associated to the periastron $p$, which is the minimum distance between the photon trajectory and the WH, see Fig. \ref{fig:Fig1}. This value depends both on the emission radius $R$, and the emission angle $\alpha_E$, and can be calculated by imposing $dr/d\psi=0$ (i.e., considering the integrating function of Eq. (\ref{eq:LB}) equal to zero),   \cite{Misner1973,Chandrasekhar1983,Defalco2016}
\begin{equation} \label{eq:p}
p^2-b_{\rm ph}^2e^{2\Phi(p)}=0,
\end{equation}
with the further condition that $p\ge r_{ps}$. Therefore, we can calculate the periastron bending angle $\psi_p=\psi_p(p,\alpha_p)$\footnote{It is important to stress that $\psi_p$ is an angle that depends on the selected photon trajectory, because the value of $p$ depends on the impact parameter $b_{\rm ph}$ (or equivalently on the emission angle $\alpha_E$) and the emission radius $R$, see Fig. \ref{fig:Fig1}. }. Using Eq. (\ref{eq:LB}), we calculate $\psi_S$, and employing another symmetrization process with respect to the periastron bending angle $\psi_p$, we finally determine the desired bending angle $\psi_E=2\psi_p-\psi_S$, associated to $\alpha_E$.

\subsubsection{Time delay}
\label{sec:TD}
A photon following its null geodesic in the metric (\ref{eq:MTmetric}) from a generic emission point at distance $R$ to the observer location at infinity will take an infinite travel time. To avoid such singularity, we consider the relative time delay between a photon emitted at radius $R$ with a generic impact parameter $b_{\rm ph}$, $T(b_{\rm ph})$, and another photon emitted always at distance $R$ (but of course at different azimuthal position $\varphi$) with radial impact parameter, i.e., $b_{\rm ph}=0$, see Fig. \ref{fig:Fig1}. Therefore, the non-singular photon travel time delay reads as \cite{Pechenick1983,Poutanen2006} 
\begin{eqnarray} \label{eq:TD}
\Delta t(b_{\rm ph})&&=\int_{t_0}^{\bar{t}} dT(b_{\rm ph})-dT(0)\\
&&=\int_R^\infty\left[\frac{dT(b_{\rm ph})}{dr}-\frac{dT(0)}{dr}\right]dr\nonumber\\
&&=\int_R^\infty\left[\frac{k^t}{k^r}\Big|_{b_{\rm ph}}-\frac{k^t}{k^r}\Big|_{b_{\rm ph}=0}\right]dr=\nonumber\\
&&=\int_R^\infty\frac{e^{-\Phi(r)}}{\sqrt{1-b(r)/r}}\left\{\left[1-\frac{b_{\rm ph}^2}{r^2}e^{2\Phi(r)}\right]^{-1/2}-1\right\}dr.\nonumber
\end{eqnarray}
This formula is valid for $0\le\alpha_E\le\pi/2$. However, if we have photons endowed with turning points $\pi/2\le\alpha_E\le\alpha_{\rm max}$ and an impact parameter $b_{\rm ph}\ge b_c$, we make use of the periastron $p$, see Eq. (\ref{eq:p}). Indeed, we split the calculation of the total time delay in the following three parts along the photon trajectory (see Fig. \ref{fig:Fig1}): 
\begin{itemize}
\item time delay between $[\pi/2,\alpha_E]$, given by\footnote{We prefer to use the module notation to calculate this time, because a-priori we do not know whether $p$ is greater or lower than $R$, and a time cannot be negative.}
\begin{equation}
\Delta t_{E-p}(b_{\rm ph})=\left|\int_R^p\frac{e^{-\Phi(r)}}{\sqrt{1-b(r)/r}}\left[1-\frac{b_{\rm ph}^2}{r^2}e^{2\Phi(r)}\right]^{-1/2}dr\right|;
\end{equation}
\item then using the symmetrization process we determine $\alpha_S=\pi-\alpha_E\in[0,\pi/2]$. Since the integrating function of Eq. (\ref{eq:TD}) is symmetric with respect to $\alpha_p=\pi/2$, we have that $\Delta t_{p-S}=\Delta t_{E-p}$;
\item finally we consider the last part of the orbit, which gives the time delay from $\alpha_S$ to the observer at infinity, $\Delta t_S$ (as that of a photon with zero-turning points), which can be calculated through Eq. (\ref{eq:TD}).
\end{itemize}
Summing up these three pieces, we obtain the proper formula of the travel time delay with turning points
\begin{equation}
\Delta t(b_{\rm ph})=\Delta t_S(b_{\rm ph})+2\Delta t_{E-p}(b_{\rm ph}).
\end{equation}

\subsubsection{Solid angle}
\label{sec:SA}
We consider the emission reference frame $\left\{\boldsymbol{x},\boldsymbol{y},\boldsymbol{z}\right\}$ and the observer reference frame $\left\{\boldsymbol{x'},\boldsymbol{y'},\boldsymbol{z'}\right\}$, where the two systems are rotated by the observer inclination angle $i$ around $\boldsymbol{y}=\boldsymbol{y'}$. We use spherical coordinates in both systems, and we differentiates them by labelling those belonging to the observer reference frame by primes. In this case, it is more convenient to use the following spherical coordinates $\left\{r,\psi,\varphi\right\}$ in the emission reference frame $\left\{\boldsymbol{x},\boldsymbol{y},\boldsymbol{z}\right\}$, where $r$ is the radius, $\psi$ is the polar angle measured from the $\boldsymbol{z}$-axis, and $\varphi$ is the azimuthal angle measured clockwise from the positive $\boldsymbol{x}$-axis in the $\boldsymbol{x}$ -- $\boldsymbol{y}$ plane. The same definitions hold also for the spherical coordinates $\left\{r',\psi',\varphi'\right\}$ in the observer reference frame $\left\{\boldsymbol{x'},\boldsymbol{y'},\boldsymbol{z'}\right\}$ (see Fig. \ref{fig:Fig1}, for a visual representation).

The solid angle, $d\Omega$, in the observer reference frame reads as $d\Omega=\sin\psi' \,d\varphi' \, d\psi'$. The observer is located at a distance $D$ far from the emission point $R$, i.e., $R\ll D<\infty$, which can be considered as at infinity. The solid angle equation can be also expressed in terms of the impact parameter, $b_{\rm ph}$, by its first-order approximation for infinitesimally small $\psi'$ (i.e., $b_{\rm ph}\approx D\ \psi'$) as \cite{Defalco2016,Defalco2020WH}
\begin{equation} 
\label{CASF} 
d\Omega=\frac{b_{\rm ph}\ db_{\rm ph}\ d\varphi'}{D^2}.
\end{equation}
In the emission reference frame, Eq. (\ref{CASF}) becomes
\begin{equation} \label{NCASF} 
d\Omega=\frac{b_{\rm ph}}{D^2}\frac{\partial \varphi'}{\partial \varphi}\frac{\partial b_{\rm ph}}{\partial r} dR\ d\varphi,
\end{equation}
where we considered the following dependencies $\varphi=\varphi(\varphi')$ and $b_{\rm ph}=b_{\rm ph}(r,\psi)$. The Jacobian of the transformation is always $\frac{\partial \varphi'}{\partial \varphi}\frac{\partial b_{\rm ph}}{\partial r}$ independent of the value of $\frac{\partial b_{\rm ph}}{\partial \psi}$, since the photon moves in an invariant plane. Therefore, Eq. (\ref{NCASF}) is valid for any emission point \cite{Defalco2016}. Using the following coordinates transformation $\cos\psi=\sin i\cos\varphi$ that relates the angles in the observer and emission reference frames, the term $\frac{\partial b_{\rm ph}}{\partial r}=-\frac{\partial b_{\rm ph}}{\partial \psi}\frac{\partial \psi}{\partial r}$ is calculated through the light bending Eq. (\ref{eq:LB}), and the solid angle equation reads as \cite{Defalco2016}
\begin{equation} \label{eq:SA} 
d\Omega=\frac{\frac{\cos i}{D^2\ e^{2\Phi(R)}\ \sin^2\psi_E}\frac{\sin^2\alpha_E}{\cos\alpha_E}} {\int_R^\infty \frac{dr}{r^2}\left\{\left[1-\frac{b(r)}{r}\right]\left[e^{-2\Phi(r)}-\frac{b_{\rm ph}^2}{r^2}\right]\right\}^{-3/2}}\ dR\ d\varphi.
\end{equation}
This formula is valid for $0\le\alpha_E\le\pi/2$. For photons presenting turning points $\pi/2\le\alpha_E\le\alpha_{\rm max}$, we follow the same symmetrization processes already explained for the light bending case, see Sec. \ref{sec:LB}. 

\subsubsection{Timelike ISCO and image at infinity}
\label{sec:ISCO}
In order to understand the WH spacetime geometry and infer useful information about it, we calculate the ISCO radius $r_{\rm ISCO}$ for a test particle. We follow an analogue procedure for deriving the photon sphere radius, see Eq. (\ref{photonsphere1}). We consider the timelike geodesic equation
\begin{equation} \label{eq:tg}
\begin{aligned}
&g_{\alpha\beta}\frac{dx^\alpha}{d\tau}\frac{dx^\beta}{d\tau}=-1,\\
&\Rightarrow\ \frac{E_p^2}{e^{2\Phi(r)}}-\frac{\dot{r}^2}{1-\frac{b(r)}{r}}-\frac{L_{d,p}^2}{r^2}=1,
\end{aligned}
\end{equation}
where $E_p,L_{d,p}$ are respectively the conserved energy and angular momentum along the test particle trajectory, and $\tau$ is the affine parameter along the test particle trajectory. Considering the effective potential $V_p(r)$ for circular orbits $\dot{r}=0$, we obtain
\begin{equation} \label{eq:Vp}
E_p^2\equiv V_p(r)=e^{2\Phi(r)}\left(\frac{L_{d,p}^2}{r^2}+1\right).    
\end{equation}
Imposing $[dV_p(r)/dr]_{r=r_{\rm ISCO}}=0$, we obtain
\begin{equation} \label{eq:D2}
L^2_{d,p}[\Phi'(r)r-1]+\Phi'(r)r^3=0,    
\end{equation}
which solved for the lowest value of $L_{d,p}$, permits to determine the value of the timelike ISCO radius $r_{\rm ISCO}$.

An emitting source around a WH can be ray-traced from the emission location to the observer screen using the following coordinates (see Ref. \cite{Cunningham1973,Luminet1979}, for details)
\begin{equation}
\begin{cases}    
x'=-b_{\rm ph}\sin\varphi'\equiv-b_{\rm ph}\frac{\sin\varphi}{\sin\psi_E},\\
\\
y'=-b_{\rm ph}\cos\varphi'\equiv-b_{\rm ph}\frac{\cos i\cos\varphi}{\sin\psi_E}.
\end{cases}
\end{equation}

\section{General relativistic Poynting-Robertson effect in Morris-Thorne-like wormhole spacetimes}
\label{sec:PRE}
In a previous work \cite{Defalco2020WH} we derived the equations of motion of a test particle moving in the gravitational field of a WH and influenced also by a radiation field, including the general relativistic PR effect, from a spherical and rigidly rotating emitting source. The metric (\ref{eq:MTmetric}) can be written in the emission reference frame $\left\{\boldsymbol{x},\boldsymbol{y},\boldsymbol{z}\right\}$, see Sec. \ref{sec:SA}, by employing spherical coordinates $\{r,\theta,\varphi \}$. In this coordinate system we have
\begin{equation} \label{eq:MTmetric2}
ds^2=-e^{2\Phi(r)}dt^2+\frac{dr^2}{1-b(r)/r}+r^2(d\theta^2+\sin^2\theta d\varphi^2).
\end{equation}
The radiation field is modelled as a coherent flux of photons, travelling along null geodesics on the background spacetime (\ref{eq:MTmetric2}), and endowed with an impact parameter $\lambda$, whose expression is 
\begin{equation} \label{MT_impact_parameter}
\lambda=\Omega_{\star}\left[\frac{\mathrm{g_{\varphi\varphi}}}{-\mathrm{g_{tt}}}\right]_{r=R_\star}, 
\end{equation}
where $R_\star$ and $\Omega_\star$ are, respectively, the radius and angular velocity of the emitting surface. The lower label in the right square bracket indicates that the metric components $g_{\varphi\varphi},g_{tt}$ must be evaluated at $r=R_\star$. It is important to note that $\lambda$ is different from that of Eq. (\ref{eq:bim}).

The local static observer (SO) frames $\left\{\boldsymbol{e_{\hat t}},\boldsymbol{e_{\hat r}},\boldsymbol{e_{\hat \varphi}},\boldsymbol{e_{\hat \psi}}\right\}$ are our family of fiducial observers. In the SO frames, we can measure clockwise from the $\boldsymbol{e_{\hat \varphi}}$-axis the emission angle $\beta$, related to the impact parameter $\lambda$, which is 
\begin{equation} \label{SO_angle}
\cos\beta=\frac{e^{\Phi(r)}}{r}\lambda. 
\end{equation}

We assume that the interaction between the radiation and the test particle  is described by Thomson scattering, which is characterized by a constant momentum-transfer cross section $\sigma$, which in turn is independent of the direction and the frequency of the radiation field. The equations of motion governing the dynamics of a test particle orbiting around a WH in the equatorial plane  and influenced by a radiation field, which also includes the general relativistic PR effect, are given by \cite{Defalco2020WH}: 
\begin{eqnarray}
\frac{d\nu}{d\tau}&=& -\frac{\sin\alpha}{\gamma}a(n)^{\hat r}\label{EoM1}\\
&&+\frac{ A [1-\nu\cos(\alpha-\beta)][\cos(\alpha-\beta)-\nu]}{e^{2\Phi(r)}\sqrt{r^2-e^{2\Phi(r)}\lambda^2}},\nonumber\\
\frac{d\alpha}{d\tau}&=&-\frac{\gamma\cos\alpha}{\nu}\left[a(n)^{\hat r}+k_{\rm (Lie)}(n)^{\hat r}\,\nu^2\right]\label{EoM2}\\
&&+\frac{ A [1-\nu\cos(\alpha-\beta)]\sin(\beta-\alpha)}{e^{2\Phi(r)}\sqrt{r^2-e^{2\Phi(r)}\lambda^2}\ \nu\cos\alpha},\nonumber\\
U^{\hat r}&\equiv&\frac{dr}{d\tau}=\frac{\gamma\nu\sin\alpha}{\sqrt{g_{rr}}}, \label{EoM3}\\
U^{\hat \varphi}&\equiv&\frac{d\varphi}{d\tau}=\frac{\gamma\nu\cos\alpha}{\sqrt{g_{\varphi\varphi}}},\label{EoM4}\\
U^{\hat t}&\equiv&\frac{dt}{d\tau}=\frac{\gamma}{e^{\Phi(r)}},\label{time}
\end{eqnarray}
where $\tau$ is the affine parameter (proper time) along the test particle trajectory, $\boldsymbol{U}$ is the test particle four-velocity, $\gamma(U,n)\equiv\gamma=1/\sqrt{1-||\boldsymbol{\nu}(U,n)||^2}$ is the Lorentz factor, $\nu=||\boldsymbol{\nu}(U,n)||$ is the magnitude of the test particle spatial velocity, and $\alpha$ is the azimuthal angle of the vector $\boldsymbol{\nu}(U,n)$ measured clockwise from the $\boldsymbol{e_{\hat \varphi}}$ direction in the SO frame. Furthermore, $A$ is the luminosity parameter, which can be equivalently written as $A/M=\mathcal{L}/\mathcal{L}_{\rm EDD}\in[0,1]$ with $\mathcal{L}$ the emitted luminosity at infinity, and $\mathcal{L}_{\rm EDD}$ the Eddington luminosity. The SO kinematical quantities $\boldsymbol{a}(n)$ and $\boldsymbol{k_{(\rm Lie)}}(n)$ denote, respectively, the acceleration and the relative Lie curvature vector, whose expressions are
\begin{eqnarray}
a(n)^{\hat r}&=&\Phi'(r)\sqrt{1-b(r)/r},\\
k_{\rm (Lie)}(n)^{\hat r}&=&-\frac{\sqrt{1-b(r)/r}}{r}.
\end{eqnarray}

\subsection{Critical hypersurfaces}
\label{sec:CH}
An important implication of the dynamical system (\ref{EoM1})--(\ref{EoM4}) is the \emph{possible} existence of a critical hypersurface, which is the region where there is a perfect balance between the radiation and gravitational forces and the test particle has a perpetual stable motion. The contingency of the critical hypersurface strictly depends on the considered metric (see discussions in Sec. II.C.2 in Ref. \cite{Defalco2020WH}, for more details). The equation defining the critical hypersurface is \cite{Defalco2020WH}
\begin{equation} \label{eq:CH}
a(n)^{\hat r}+k_{\rm (Lie)}(n)^{\hat r}\cos^2\beta-\frac{A \sin^3\beta}{r^2\ e^{2\Phi(r)}}=0.
\end{equation}
If a critical hypersurface exists, the test-particle must move on it with constant velocity $\nu=\cos\beta$. Equation (\ref{eq:CH}) may admit none, one, or multiple roots, depending on the specific WH solution. However, even in this general case, the critical hypersurface depends continuously on the luminosity parameter $A$, because it is written both in terms of the metric components (which by definition are smooth functions) and of the radiation force components (which are also smooth functions defined on the whole spacetime). This is a fundamental propriety which allows the formation of such structures in regions close to the WH throat, where  strong  gravitational fields occur. Therefore, the existence of PR critical hypersurface provides essential information both on the WH geometry and on those extended theories of gravity which turn out to be more suited to describe their emission properties in strong gravitational regimes \cite{Defalco2020WH}. 

\section{Radiation flux emitted by a source around a spherically symmetric and static wormhole}
\label{sec:flux}
In this geometrical environment, see Sec. \ref{sec:RTE}, we consider as emission point a test particle driven by Eqs. (\ref{EoM1}) -- (\ref{time}), and having coordinates $P=(r,\varphi)$. If there exists the PR critical hypersurface, which is a circular orbit in the equatorial plane, then we have that $r=r_{\rm crit}$ is determined by solving Eq. (\ref{eq:CH}), and $\varphi=\omega_{PR}t$. At $t=0$  the emission point is closest to the observer direction, and 
\begin{equation}
\omega_{PR}=\frac{\cos\beta}{r_{\rm crit}},
\end{equation}
is the PR angular velocity measured by the observer at infinity. The photon arrival time $T_{\rm obs}$ is the sum of the emission time $T_{\rm em}=\varphi/\omega_k$ and the photon propagation delay $\Delta t(b_{ph})$, i.e., $T_{\rm obs}=T_{\rm em}+\Delta t(b_{ph})$ (see Eq. (\ref{eq:TD})). 

The observed flux at the emission frequency $\nu_{\rm em}$ is \cite{Defalco2016}
\begin{equation} \label{eq:flux}
F_{\nu_{\rm em}}=\int_{\Omega}\frac{\epsilon_0 \xi^q}{4\pi}(1+z)^{-4}d\Omega,
\end{equation}
where $\epsilon_0$ is the vacuum permittivity, $\xi=R/M$  the local surface emissivity, $q\in\mathbb{R}$ the emission index, and $(1+z)^{-1}=E_{\rm obs}/E_{\rm em}$ the gravitational redshift, i.e., the ratio between the observed $E_{\rm obs}$ and emitted $E_{\rm em}$ energies.

\subsection{Gravitational redshift}
Now, we have all the elements to calculate the gravitational redshift $(1+z)^{-1}$ associated to the motion of the test particle. When the PR critical hypersurface exists, the test particle moves on it with velocity
\begin{equation}
U^\alpha=\frac{1}{|\sin\beta|}\left(\frac{1}{e^{\Phi(r_{\rm crit})}},0,0,\frac{\cos\beta}{r_{\rm crit}}\right).
\end{equation}
We transform this velocity from the emission reference frame $\left\{\boldsymbol{x},\boldsymbol{y},\boldsymbol{z}\right\}$ to the observer reference frame $\left\{\boldsymbol{x'},\boldsymbol{y'},\boldsymbol{z'}\right\}$ (see Sec. \ref{sec:SA} and \cite{Misner1973,Defalco2016}),
\begin{eqnarray}
U^{t'}&=&U^t,\qquad U^{r'}=U^r,\\
U^{\theta'}&=& \frac{\partial\theta'}{\partial\varphi}U^\varphi+\frac{\partial\theta'}{\partial\theta}U^\theta,\quad
U^{\varphi'}= \frac{\partial\varphi'}{\partial\varphi}U^\varphi+\frac{\partial\varphi'}{\partial\theta}U^\theta.\notag
\end{eqnarray}
Since $U^\theta=0$ and the transformations between these two reference frames are \cite{Defalco2016}
\begin{equation}
\sin\varphi=\sin\theta'\sin\varphi',\quad \cos\theta'=\sin i\cos\varphi,    
\end{equation}
then the test particle velocity in the observer frame reads as
\begin{equation}
U^{\alpha'}=\frac{1}{|\sin\beta|}\left(\frac{1}{e^{\Phi(r_{\rm crit})}},0,\frac{\cos\beta}{r_{\rm crit}}\frac{\sin i\sin\varphi}{\sin\psi},U^{\varphi'}\right),
\end{equation}
where the explicit expression of  $U^\varphi{}'$ is not necessary for the calculation of the gravitational redshift, as it will be clearer soon. We used also that $\theta'\equiv\psi$, corresponding to the light bending angle, see Sec. \ref{sec:LB} and Fig, \ref{fig:Fig1}.

The photon velocity is (see Sec. \ref{sec:RTE}, for details) 
\begin{equation}
k_\alpha=\left(-1,-\sqrt{\frac{e^{-2\Phi(R)}-\frac{\lambda^2}{R^2}}{1-\frac{b(R)}{R}}},-b_{\rm ph},0\right).
\end{equation}

Since the observer is static, its velocity is \begin{equation}
V^\alpha=(1,0,0,0).
\end{equation}

Since $E_{\rm em}=(U^\alpha k_\alpha)_{\rm em}$ and $E_{\rm obs}=(V^\alpha k_\alpha)_{\rm obs}$, the gravitational redshift is \cite{Misner1973,Defalco2016}
\begin{equation} \label{eq:redshift1}
\begin{aligned}
(1+z)^{-1}&\equiv\frac{E_{\rm obs}}{E_{\rm em}}\\
&=\frac{|\sin\beta|r_{\rm crit}e^{\Phi(r_{\rm crit})}}{r_{\rm crit}+b_{\rm ph}\cos\beta e^{\Phi(r_{\rm crit})}\frac{\sin i\sin\varphi}{\sin\psi}}.
\end{aligned}
\end{equation}

\subsubsection{Emission from a disk}
Let us consider an accretion disk lying in the equatorial plane $\theta=\pi/2$ around a WH and extending from $R_{\rm in}$ to $R_{\rm out}$. We assume that the matter inside the disk moves on circular orbits with the Keplerian angular velocity $\Omega_K$, which has a proper expression for each WH metric. In order to estimate the gravitational redshift, we first need to derive  the expression of $\Omega_K$. 

We consider the timelike geodesic equation (\ref{eq:tg}) for stable circular orbits (i.e., $\dot{r}=0$ and $dV_p/dr=0$). From Eq. (\ref{eq:D2}) we obtain the test particle's angular momentum,
\begin{equation} \label{eq:Lzp}
L_{d,p}^2=\frac{r^3\Phi'(r)}{1-r\Phi'(r)}.    
\end{equation}
Substituting this expression in Eq. (\ref{eq:Vp}), we derive  the expression of the test particle's energy
\begin{equation} \label{eq:Ep}
E_p^2=\frac{e^{2\Phi(r)}}{1-r\Phi'(r)}.    
\end{equation}
The two explicit expressions (\ref{eq:Lzp}) and (\ref{eq:Ep}) of the test particle's constants of motion permit to obtain the test particle's velocity components in the following way:
\begin{eqnarray}
U^{\hat t}&\equiv&g^{tt}U_t=\frac{1}{e^{\Phi(r)}\sqrt{1-r\Phi'(r)}},\\
U^{\hat\varphi}&\equiv&g^{\psi\psi}U_\psi=\sqrt{\frac{\Phi'(r)}{r(1-r\Phi'(r))}}.
\end{eqnarray}
Therefore, the Keplerian angular velocity $\Omega_K$ for the WH metric (\ref{eq:MTmetric}) is given by
\begin{equation} \label{eq:OmK}
\Omega_K\equiv\frac{U^{\hat\varphi}}{U^{\hat t}}=e^{\Phi(r)}\sqrt{\frac{\Phi'(r)}{r}}.   
\end{equation}
A test particle moving on a circular orbit has the following velocity field  \cite{Misner1973,Defalco2016}
\begin{equation}
\begin{aligned}
U^{\hat\alpha}&=U^{\hat t}(1,0,0,\Omega_K\sin\varphi')\\
&=U^{\hat t}\left(1,0,0,\Omega_K\frac{\sin i\sin\varphi}{\sin\psi}\right),
\end{aligned}
\end{equation}
where we have used the transformation between the observer frame and the emission frame \cite{Defalco2016}. 

The gravitational redshift for the accretion disk is \cite{Defalco2016}
\begin{equation}\label{eq:reddisk}
(1+z)^{-1}=\frac{e^{\Phi(r)}\sqrt{1-r\Phi'(r)}}{1+b_{\rm ph}\Omega_K\frac{\sin i\sin\varphi}{\sin\psi}}.    
\end{equation}

\section{Examples of ray-tracing applications to some WH solutions}
\label{sec:examples}
In this section, we apply the general ray-tracing method of Sec. \ref{sec:geometry} to different WH solutions belonging to distinct extended theories of gravity. We produce lightcurves and spectra of a PR critical hypersurface. In order to discuss and compare the results among the different metrics, we pose such ring at the same distance from the origin of the coordinate system in all WH spacetimes. The PR critical hypersurfaces are defined in terms of two parameters $(A,\lambda)$, see Eq. (\ref{eq:CH}). These cannot be chosen arbitrarily, on the contrary they should range in some astrophysically realistic intervals (see Table I and discussions in Ref. \cite{Defalco2020WH}). In addition, we produce also the image of an accretion disk around a WH, as seen by an observer at infinity. This step is important to better inquire the WH spacetime geometries. 

The WH literature encompasses a plethora of solutions, however we focus our attention only on those fulfilling the following requirements: geometrical proprieties of a standard WH solution as described in Sec. \ref{sec:geo_pro_WH} (in particular the asymptotic flatness, i.e., $b(r)/r\to0$ and $\Phi(r)\to0$ as $r\to\infty$\footnote{We have highlighted this propriety, since it is not trivial at all. Indeed, in the literature there are several WH solutions, which are not explicitly asymptotically flat. However, this property is recovered when the solution is smoothly matched with the Schwarzschild metric at a certain radius.}), and a non constant-redshift function (i.e., $\Phi'(r)\neq0$ for all $r\in[b_0,\infty]$). The chosen WH solutions belong to the following theories: classical GR (see Sec. \ref{sec:GR}), metric (see Sec. \ref{sec:METRIC2}), metric-affine (see Sec. \ref{sec:AFFINE-METRIC}), and teleparallel (see Sec. \ref{sec:TELEPARALLEL}). 
We underline that this section simply represents an application of the method outlined in Sec. \ref{sec:geometry} and it has the only aim of showing the potentialities and limits of the PR effect in distinguishing the WH features among different theories.

Each WH solution generally depends on some free parameters, which naturally stem out from the underlying theory. In order to fix them (being an essential step for the development of the forthcoming  numerical simulations), we impose some external constraints. First of all, we require that $r_{\rm ISCO}>r_{\rm ps}>b_0$, where $r_{\rm ISCO}$ and $r_{\rm ps}$ are calculated by employing  Eqs. (\ref{photonsphere1}) and (\ref{eq:D2}), respectively. Then, since we are interested in the study of BH mimickers, we impose that $r_{\rm ISCO}$ and $r_{\rm ps}$ coincide with those of the Schwarzschild BH metric (being $3M$ and $6M$, respectively, see Table \ref{tab:Table1}). We also recall that the Schwarzschild BH solution is described by the following redshift and shape functions \cite{Misner1973}:
\begin{equation} \label{eq:CGR}
\Phi(r)=\frac{1}{2}\log\left(1-\frac{2M}{r}\right),\quad b(r)=2M.
\end{equation}

\subsection{The case of General Relativity}
\label{sec:GR}
Within the GR framework, we examine a WH sourced by Casimir energy density and pressure and whose metric satisfies the semiclassical Einstein field equations (see Ref. \cite{Garattini2019}, for details). The redshift and shape functions are given by, respectively, \cite{Garattini2019}
\begin{equation} \label{eq:GR}
\begin{aligned}
\Phi(r)&=\frac{1}{2}(\omega-1)\log\left(\frac{r\omega}{r\omega+b_0}\right),\\
b(r)&=\left(1-\frac{1}{\omega}\right)b_0+\frac{b_0^2}{\omega r},
\end{aligned}
\end{equation}
where the positive-definite parameter $\omega$  and the WH throat radius $b_0$ are related by the relation
\begin{equation} \label{eq:b0}
   \omega = \dfrac{90 }{\pi^3 } \left(\dfrac{ b_0 }{ \ell_{\rm P}}\right)^2,
\end{equation}
$\ell_{\rm P}$ being the Planck length. The fundamental radii $r_{\rm ps}$ and $r_{\rm ISCO}$ are given by, respectively,
\begin{equation}
r_{\rm ps}=\frac{b_0}{6}(\omega-3),\quad r_{\rm ISCO}=\frac{b_0}{3}(\omega-3).
\end{equation}
Therefore, we can obtain positive-definite radii provided that $\omega>3$, whereas $\omega=3$, which is the condition leading to Casimir stress-energy tensor, gives vanishing radii. In addition, the requirement $r_{\rm ISCO}>r_{\rm ps}>b_0$ yields $\omega>9$. In order to simplify our calculations, we have considered the peculiar case $\omega=4$, where $r_{\rm ps}$ and $r_{\rm ISCO}$ are naturally in order, but located inside the  throat. This implies that in such spacetime both timelike and null circular orbits can be stable everywhere outside the WH throat. Moreover,  Eq. (\ref{eq:b0}) permits to calculate $b_0$, which expressed in gravitational radii reads as
\begin{equation} \label{eq:b0GR}
b_0=1.3\times 10^{-38}\left(\frac{M_\odot}{M}\right) M.
\end{equation}
The periastron $p$ can be obtained by looking at the real root of the  cubic algebraic equation 
\begin{equation} \label{eq: cubic equation_1}
   64 p^3+48 p^2+4(3-16b_{\rm ph}^2)p+1=0, 
\end{equation}
jointly with the condition $ p\ge r_{\rm ps}$. The real solution is:
\begin{equation}\label{eq:per1}
\begin{aligned}
p&=\frac{1}{12}\left[-3b_0 
-\frac{8\ 3^{2/3} b_{\rm ph}^{2/3}}{(\sqrt{3}\sqrt{27b_0^2-64 b_{\rm ph}^2}-9b_0)^{1/3}}\right.\\
&\left.+2 \sqrt[3]{3}b_{\rm ph}^{2/3} (\sqrt{3}\sqrt{27b_0^2-64 b_{\rm ph}^2}-9b_0)^{1/3}
\right].
\end{aligned}
\end{equation}
The condition for which the terms underneath the square root sign are non-negative leads to the inequality $b_{\rm ph}\le\sqrt{27}b_0/8\approx0.65b_0$. Since the critical impact parameter is given by $b_c=0.66\ b_0$ (i.e., $b_{\rm ph}\ge b_c$), we obtain no real solution for the periastron $p$ and this  in turn implies that \emph{no turning points exist in such WH spacetime.}

Since for emissions occurring at distances of order $M$ the throat radius $b_0$ turns out to be extremely small, it is reasonable to consider the approximation $b_0\to0$. In this case, we can tremendously simplify the integrals underlying the gravitational effects, being even able to determine their analytical expressions, which are as follows: 
\begin{eqnarray}
\psi&=&\arctan\left(\frac{b_{\rm ph}}{\sqrt{R^2-b_{\rm ph}^2}}\right),\\
\Delta t(b_{\rm ph})&=&R-\sqrt{R^2-b_{\rm ph}^2},\\
\Delta t_{E-p}(b_{\rm ph})&=&\left|\sqrt{R^2-b_{\rm ph}^2}-\sqrt{p^2-b_{\rm ph}^2}\right|,\\
d\Omega&=&\frac{1}{\sqrt{R^2-b_{\rm ph}^2}}.
\end{eqnarray}
In this case, the critical impact parameter and the periastron are given by $b_c=0$ and $p=b_{\rm ph}$, respectively. Finally, by performing accurate numerical simulations we have found that the PR critical hypersurfaces are allowed in the following parameters range: $A\in(0,0.3],\lambda\in[0,0.8]$. A concise summary of the obtained results is reported in Table \ref{tab:Table1}. 

\subsection{Metric theories}
\label{sec:METRIC2}
In the context of metric theories of gravity, let us consider $f(R)$ gravity which is a generalization/extension of Einstein's GR, defined by a generic function, $f$, of the Ricci scalar, $R$ \cite{Capozziello2009nq}. A wide range of phenomena can be addressed by this theory by adopting different forms of $f$ \cite{Capozziello:2011et,Capozziello:2012ie}. In particular, we can take into account models like 
\begin{equation}
f(R)= R-\mu R_c \left(\dfrac{R}{R_c}\right)^l,
\end{equation}
where $\mu$, $R_c$ and $l$ are constants such that $\mu>0$, $R_c >0$ and $0<l<1$ (see Ref. \cite{Godani2020} and references therein, for more details). This kind of models are particularly relevant at infrared scales, because they potentially address problems like: clustering of structures and accelerated expansion of the Hubble flow \cite{Starobinsky2007,Capozziello:2012ie}. Specifically, instead of searching for new form of exotic matter as fundamental constituents for dark energy and dark matter, the approach is devoted to solve the dark side problem through geometry. In other words, further degrees of freedom related to the gravitational field are considered instead of choosing {\it a priori} the Hilbert-Einstein action. 

Besides these fundamental problems, $f(R)$ gravity has several spherically symmetric solutions which can be interesting both for BH and WH physics. In particular a WH solution can be \cite{Godani2020}
\begin{equation} \label{eq:MT}
\Phi(r)=-\frac{\alpha}{r},\quad b(r)=\frac{r}{e^{(r-b_0)}},
\end{equation}
where $\alpha>0$. We find that the fundamental radii $r_{\rm ps}$ and $r_{\rm ISCO}$ are given by, respectively, 
\begin{equation} \label{eq:MT_fundamental_radii}
r_{\rm ps}=\alpha, \quad r_{\rm ISCO}= 2 \alpha.
\end{equation}
In this case the model sets  no constraints on $b_0$ and hence we can easily choose $b_0=M$ and $\alpha=3M$, which permits to recover the two fundamental Schwarzschild radii. We find that $b_c=8.15M$ and $p=-3M/W(-3M/b_{\rm ph})$, where $W(x)$ is the \emph{Lambert $W$ function} (also known as \texttt{ProductLog} in computer algebra framework), which is the inverse function of $xe^x$, i.e., $W(xe^x)=x$ (see Ref. \cite{FUKUSHIMA2013}, for further details and its numerical implementation in numerical codes). Imposing the condition $p\ge3M$, which implies that the photon can reach the observer at infinity, we derive that $b_{\rm ph}\ge3e$ and the function $W(x)$ must be evaluated in the branch $W_{0}(x)$ (see Ref. \cite{FUKUSHIMA2013}, for more details). After having performed some numerical simulations, we have found that the PR critical hypersurfaces exist for $A\in(0,0.6],\ \lambda\in[0,6]$. In Table \ref{tab:Table1}, a summary of the obtained results is displayed. 

\subsection{Metric-affine theories}
\label{sec:AFFINE-METRIC}
Assuming the {\it metricity} in a theory of gravity means accepting the validity of the Equivalence Principle and then the coincidence of the causal and geodesic structures of the spacetime \cite{Capozziello:2012eu}. On the other hand, more general theories can be formulated relaxing this strong hypothesis. In this perspective, one can consider the so called metric-affine theories of gravity (i.e., the so called Palatini formulation \cite{Olmo:2011uz}),  where metric and affine connections are not related through the Levi-Civita connection.

Within metric-affine theories, particularly relevant is the hybrid metric-Palatini gravity \cite{Tamanini2013,Capozziello2015}, where a $f(R)$ term is constructed \emph{\`a la} Palatini and it is added to the usual (metric) Einstein-Hilbert Lagrangian, linear in the Ricci scalar $R$. Also in these models, it is possible to find out WH solutions, as shown in \cite{Capozziello2012}. In this case, we have the following functions:
\begin{equation} \label{eq:MAT}
\Phi(r)=\Phi_0\left(\frac{b_0}{r}\right)^{\gamma},\quad 
b(r)=b_0\left(\frac{b_0}{r}\right)^{\beta},
\end{equation}
where $\gamma>0$, $\beta >-1$, and $\Phi_0<0$. In this general framework, we obtain the photon sphere radius,
\begin{equation} \label{eq:ps-affine-metric}
     r_{\rm ps}= b_0  \left( - \Phi_0 \gamma\right)^{1/\gamma}, 
   \end{equation}
whereas the ISCO radius can be obtained by solving the following algebraic equation:  
\begin{equation} \label{eq:ISCOMAT}
    L_{d,p}^2 r^\gamma +\left(\gamma b_0^\gamma \Phi_0\right) r^2 +L_{d,p}^2 \gamma b_0^\gamma \Phi_0=0,
\end{equation}
and then minimizing the real solution in terms of $L_{d,p}$, see Sec. \ref{sec:ISCO}. The degree of Eq. (\ref{eq:ISCOMAT})
is determined once the value of $\gamma$ is known. To differentiate this metric from the previous examples, we choose $\gamma=3$, obtaining thus for the photon sphere radius
\begin{equation}\label{eq:ps-affine-metric-2}
    r_{\rm ps}= b_0  \left( - 3 \Phi_0 \right)^{1/3},
\end{equation}
while for $r_{\rm ISCO}$ we should solve the following equation in terms of the test particle angular momentum $L_{d,p}$:
\begin{equation} \label{eq:cubic_alg_equat}
r^3 + 3 b_0^3 \Phi_0 + \frac{3 r^2 b_0^3 \Phi_0}{L_{d,p}^2}=0.
\end{equation}
For simplicity, we chose $\beta=1$, $b_0=M$, and $\Phi_0=-1$. The explicit expression(s) of the real solution(s) of a cubic algebraic equation can be obtained by exploiting the Cardano formula  \cite{Abramowitz1965}. In our case, Eq. (\ref{eq:cubic_alg_equat}) admits one real solution which can written in terms of the delta function $\sqrt{3L^6_{d,p}+4M^6}$. Imposing that the quantity under the square root be non-negative, we obtain
\begin{equation}
L^6_{d,p}=-\frac{4M^6}{3}.
\end{equation}
This entails that $L_{d,p}$ should assume imaginary values, representing  a non-physical possibility. Therefore, we conclude that there are no stable circular orbits in such WH spacetime. The critical impact parameter and the periastron are respectively $b_c=2.01M$ and $p=-\sqrt[3]{3}M/\sqrt[3]{W\left[-3(M/b_{\rm ph})^3\right]}$, which implies $b_{\rm ph}\ge\sqrt[3]{3e}$ in order to have meaningful results (see discussions of Sec. \ref{sec:METRIC2}). After having performed some numerical simulations, we have found that
the PR critical hypersurfaces exist for $A\in(0,0.56],\ \lambda\in[0,2]$. In Table \ref{tab:Table1}, we report a short summary of the obtained results. 

We note that \emph{the PR effect configures as one of the viable mechanisms allowing for the occurrence of timelike stable circular orbits in such WH spacetime}.

\subsection{Teleparallel theories}
\label{sec:TELEPARALLEL}

As a final example, let us consider teleparallel equivalent gravity and its possible generalizations \cite{Cai:2015emx}. Here the Lagrangian of the gravitational field is a function of the torsion scalar $T$ and instead of using the Levi-Civita connection, the Weitzenb\"{o}ck connection is adopted. It is important to stress that the teleparallel formulation of gravity is equivalent to GR, because the field equations written in terms of $T$ can be reconciled to those written in terms of $R$. Nevertheless it is important to note that two main differences exists between $f(T)$ and $f(R)$ gravity theories: (1) in the metric formulation of $f(R)$, the field equations are of fourth-order, while in $f(T)$, they remain of second-order as it also occurs in GR; (2) in teleparallel gravity the dynamical variables are the tetrad fields, while in metric formulation, the dynamics is related to tensor metric $g_{\mu\nu}$.

In particular, we can consider a (modified) teleparallel model as $f(T,T_{\mathcal{G}})$ gravity, where $T$ is the torsion scalar and $T_{\mathcal{G}}$ is the teleparallel equivalent of the Gauss-Bonnet term (see Ref. \cite{Kofinas2014,  Bahamonde:2016jqq, Capozziello:2016eaz}, for more details). 

Also in this case, it is possible to build up WH solutions. Considering the model
\begin{equation}
    f(T,T_{\mathcal{G}}) = a_0\sqrt{a_1\, T_{\mathcal{G}}+T^2}-T,
\end{equation}
where $a_0$ and $a_1$ are dimensionless coupling constants \cite{Kofinas2014b}, a WH solution is \cite{Sharif2018} 
\begin{equation} \label{eq:TelM}
\Phi(r)=-\frac{\xi}{r},\quad b(r)=\frac{b_0^{\zeta+1}}{r^\zeta},
\end{equation}
where $\xi>0$ and $\zeta>-1$. The fundamental radii assume the same form as those of the metric theory (cf. Eqs. (\ref{eq:MT}) and (\ref{eq:MT_fundamental_radii})), i.e., 
\begin{equation}
r_{\rm ps}= \xi,\quad r_{\rm ISCO}= 2 \xi,
\end{equation}
and hence we choose $b_0=M$ and $\xi=3M$, and to ease the calculations $\zeta=2$. The critical impact parameter is $b_c=8.15M$, whereas the periastron is $p=-3M/W(-3M/b_{\rm ph})$, with $b_{\rm ph}\ge 3e$ and $W(x)$ evaluated in the branch $W_{0}(x)$ (see Sec. \ref{sec:METRIC2}). After some numerical simulations we have obtained that the PR critical hypersurfaces exist for $A\in(0,0.6],\ \lambda\in[0,6]$. In Table \ref{tab:Table1}, we report a brief summary of the obtained results.
\renewcommand{\arraystretch}{1.8}
\begin{table*}[t!]
\begin{center}
\caption{\label{tab:Table1} Summary of important information related to the different WH solutions presented in this paper. The photon sphere radius $r_{\rm ps}$ and ISCO radius $r_{\rm ISCO}$  have been calculated through Eqs.   (\ref{photonsphere1}) and (\ref{eq:D2}), respectively. The lightgray cells refer to the Schwarzschild BH solution, while the others to the WH solutions.\\}	
\normalsize
\vspace{0.3cm}
\begin{tabular}{|@{} c @{}|@{} c @{}|@{} c @{}|@{} c @{}|@{} c @{}|@{} c @{}|@{} c @{}|} 
\ChangeRT{1pt}
\ \ {\bf Theory}\ \ &\ \ {\bf WH Eq.}\ \ &\ \ {\bf Critical Hypersurface}\ \ &\ \ {\bf WH throat} \ \ &\ \ {\bf Photon sphere}\ \ &\ \ {\bf ISCO}\ \ &\ \ {\bf Ref.}\ \ \\
%BH
\rowcolor{lightgray}& \ \ {\bf BH Eq.}\ \ & &\ \ {\bf BH horizon} \ \ & & & \\
\ChangeRT{1pt}
\rowcolor{lightgray}\ \  Schwarzschild\ \ & (\ref{eq:CGR}) & $A\in(0,0.6],\ \ \lambda\in[0,6]$ & $2M$ & $3M$ & $6M$ & \cite{Misner1973}\\
\hline
%GR
GR & (\ref{eq:GR}) & $A\in(0,0.3],\ \ \lambda\in[0,0.8]$ & $b_0$\footnote{For the WH in GR theory it is possible to determine the WH throat within the theory itself as $b_0=1.3\times10^{-38}\left(\frac{M_\odot}{M}\right)M$, see Eq. (\ref{eq:b0GR}).} & $\frac{b_0}{6}$ & $\frac{b_0}{3}$ & \cite{Garattini2019}\\
\hline
%METRICO
Metric & (\ref{eq:MT}) & $A\in(0,0.6],\ \ \lambda\in[0,6]$ & $M$ & $3M$ & $6M$ & \cite{Godani2020}\\
\hline
%METRICO-AFFINE
Affine-Metric & (\ref{eq:MAT}) & $A\in(0,0.56],\ \ \lambda\in[0,2]$ & $M$ & $\sqrt[3]{3}M$ &\ \ NOT EXIST\ \ & \cite{Capozziello2012}\\
\hline
%TELEPARALLELO
Teleparallel & (\ref{eq:TelM}) & $A\in(0,0.6],\ \ \lambda\in[0,6]$ & $M$ & $3M$ & $6M$ & \cite{Sharif2018}\\
\ChangeRT{1pt}
\end{tabular}
\end{center}
\end{table*}

\begin{figure*}[th!]
\centering
\hbox{
\includegraphics[scale=0.32]{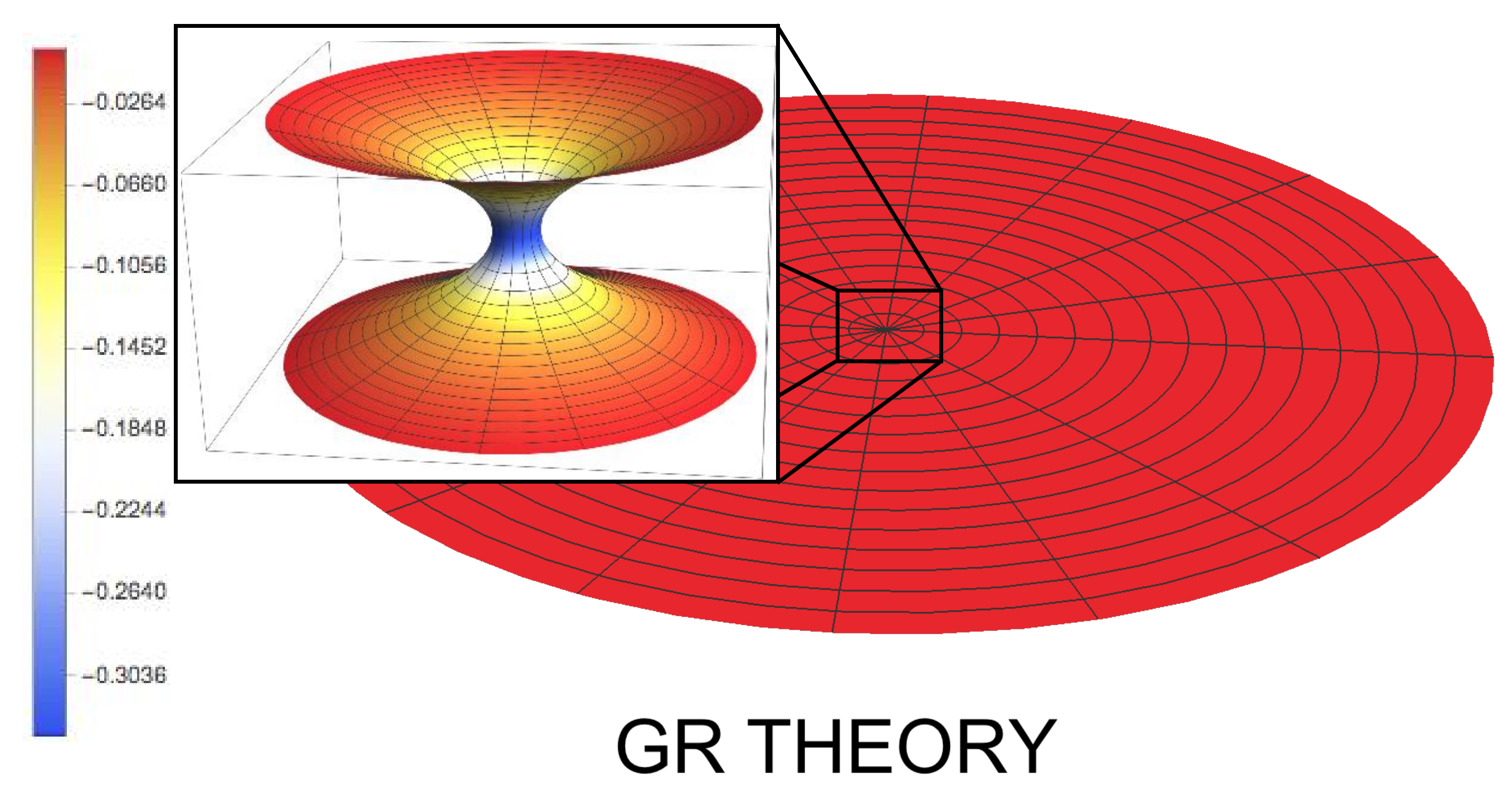}
\hspace{0.5cm}
\includegraphics[scale=0.55]{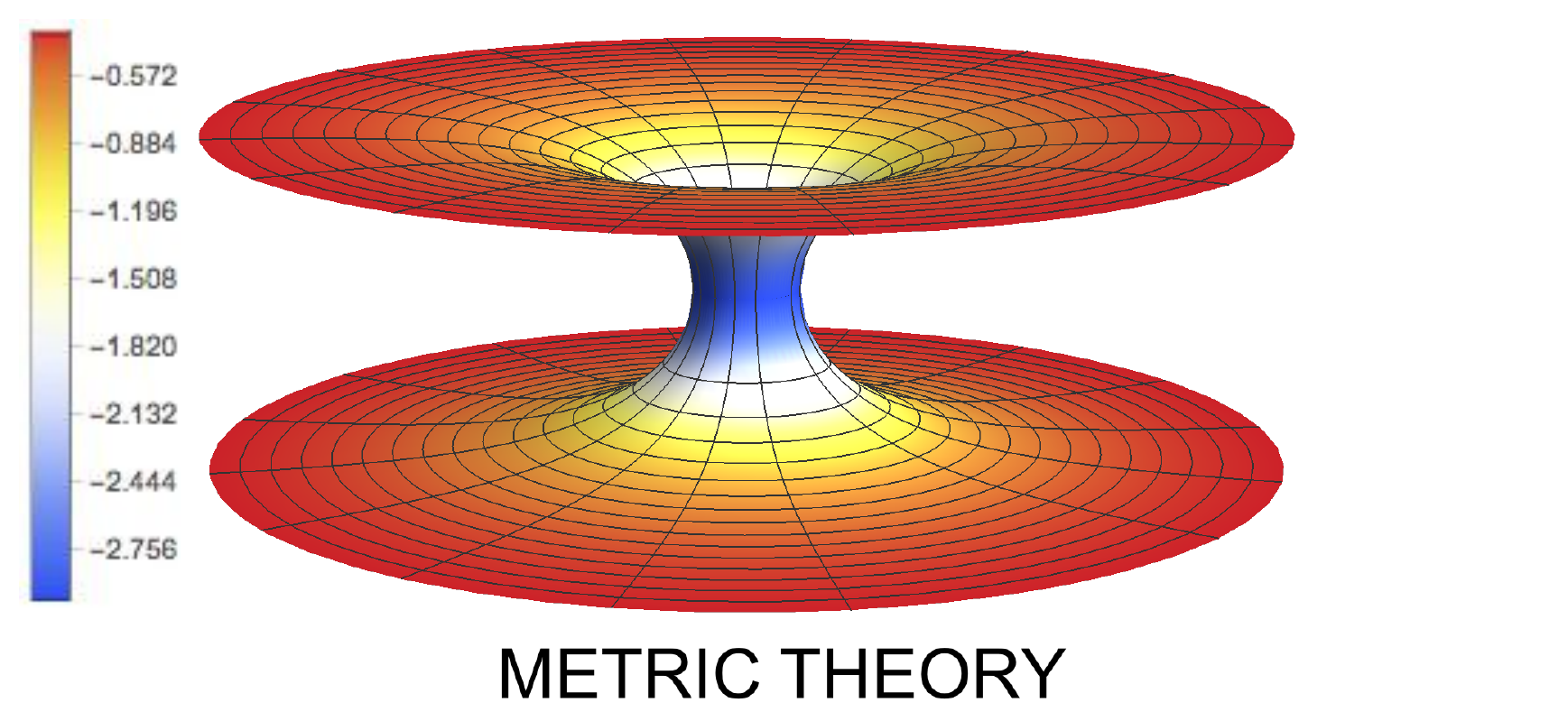}}
\vspace{0.1cm}
\hbox{\includegraphics[scale=0.55]{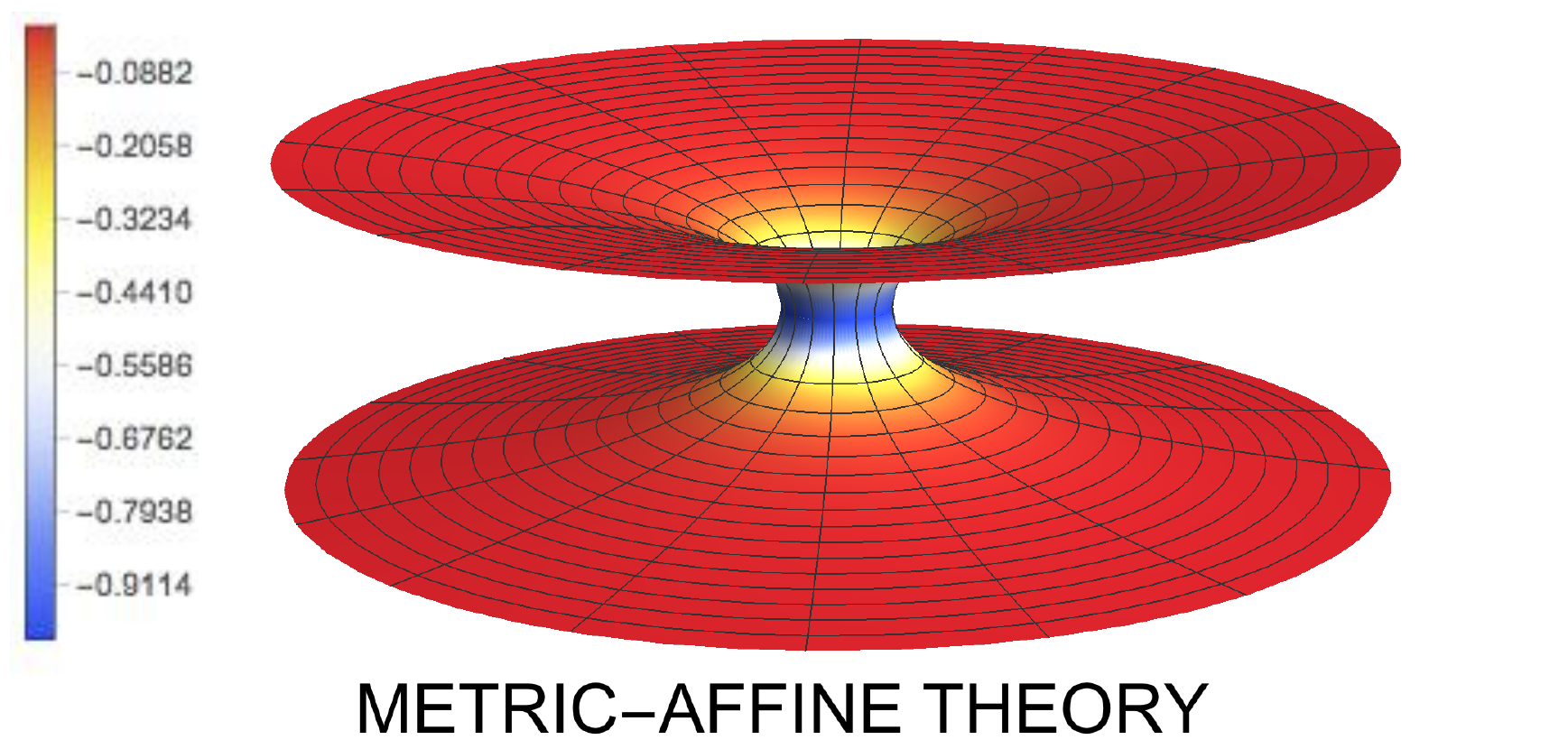}
\hspace{-0.5cm}
\includegraphics[scale=0.55]{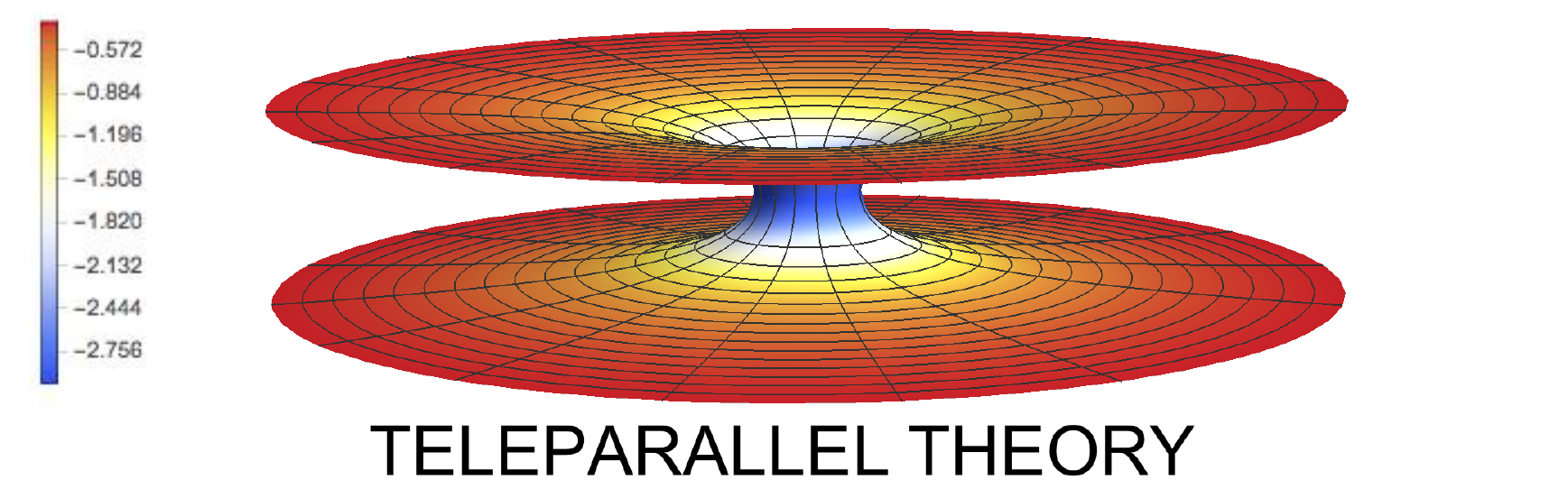}
}
\caption{Shapes of the WH solutions on which have been applied the related redshift functions. They are all plotted in $M$ units. The inset figure in GR theory represents the WH shape plotted in $b_0$ unity with the same legend bar of the main figure.}
\label{fig:Fig2}
\end{figure*}

\subsection{Numerical simulations and discussions}
\label{sec:discuss}
\begin{figure*}
\centering
%1) Location Radii
\hspace{-0.3cm}\hbox{
\includegraphics[scale=0.225]{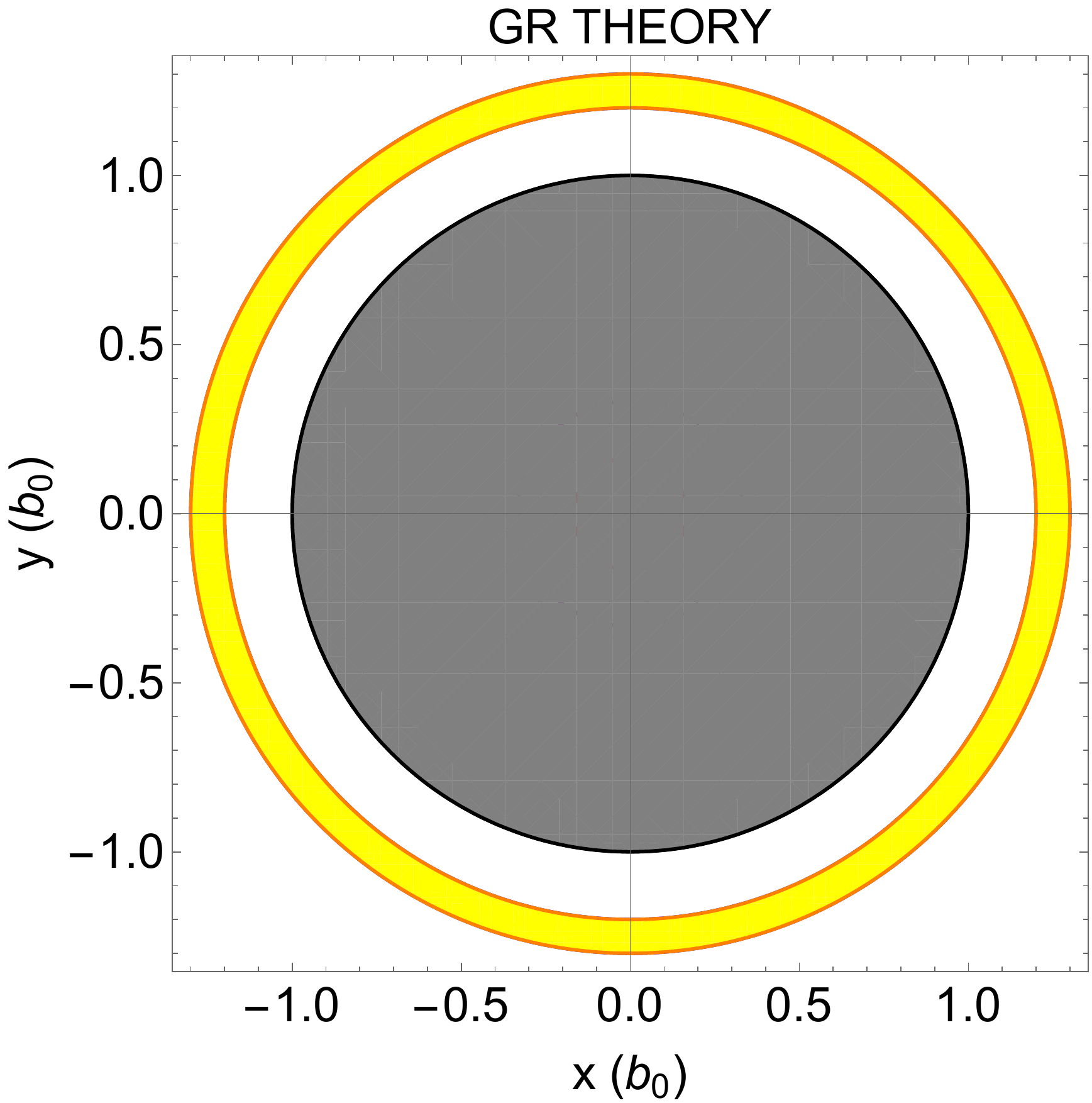}
\hspace{0.7cm}
\includegraphics[scale=0.21]{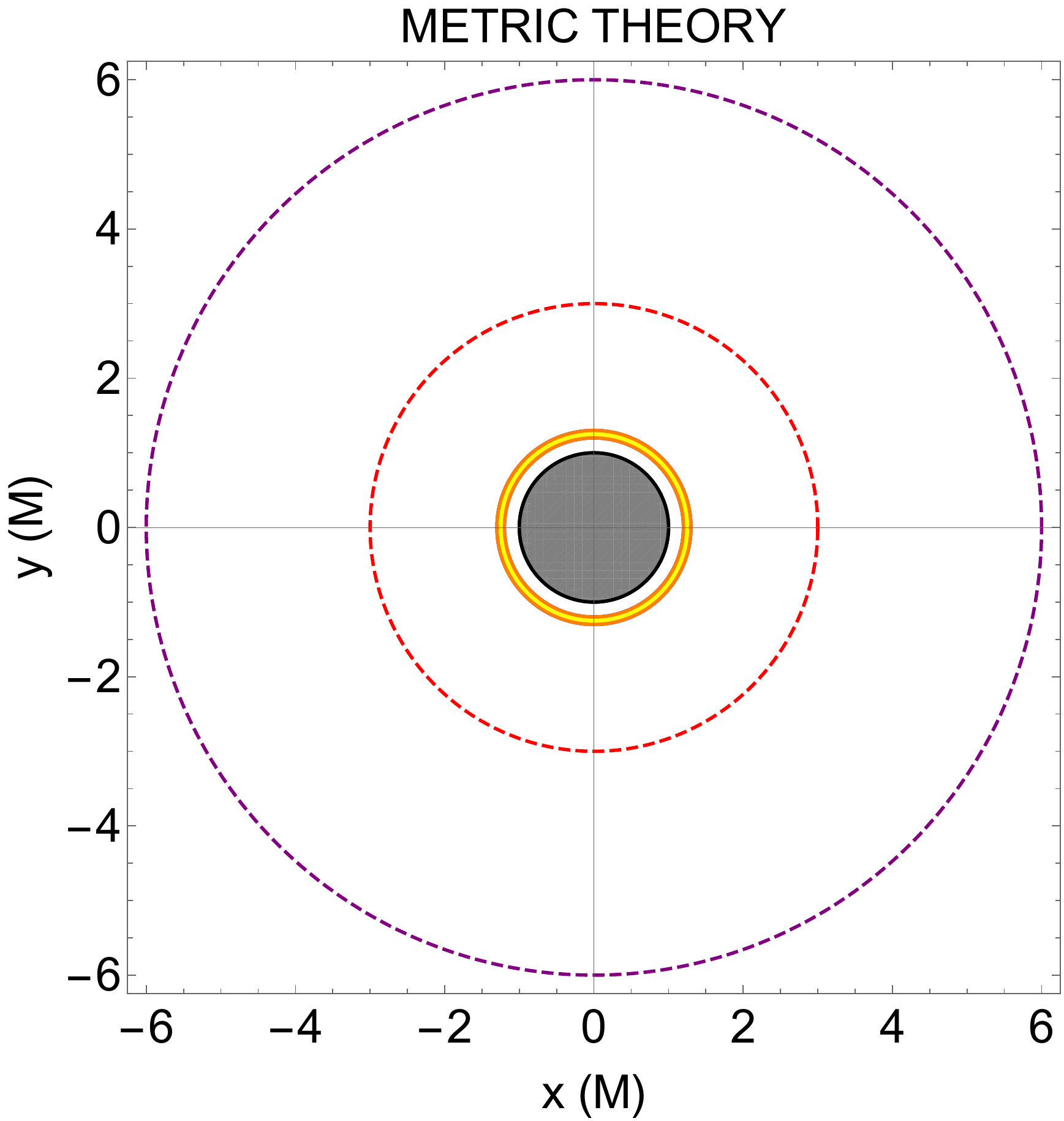}
\hspace{0.6cm}
\includegraphics[scale=0.228]{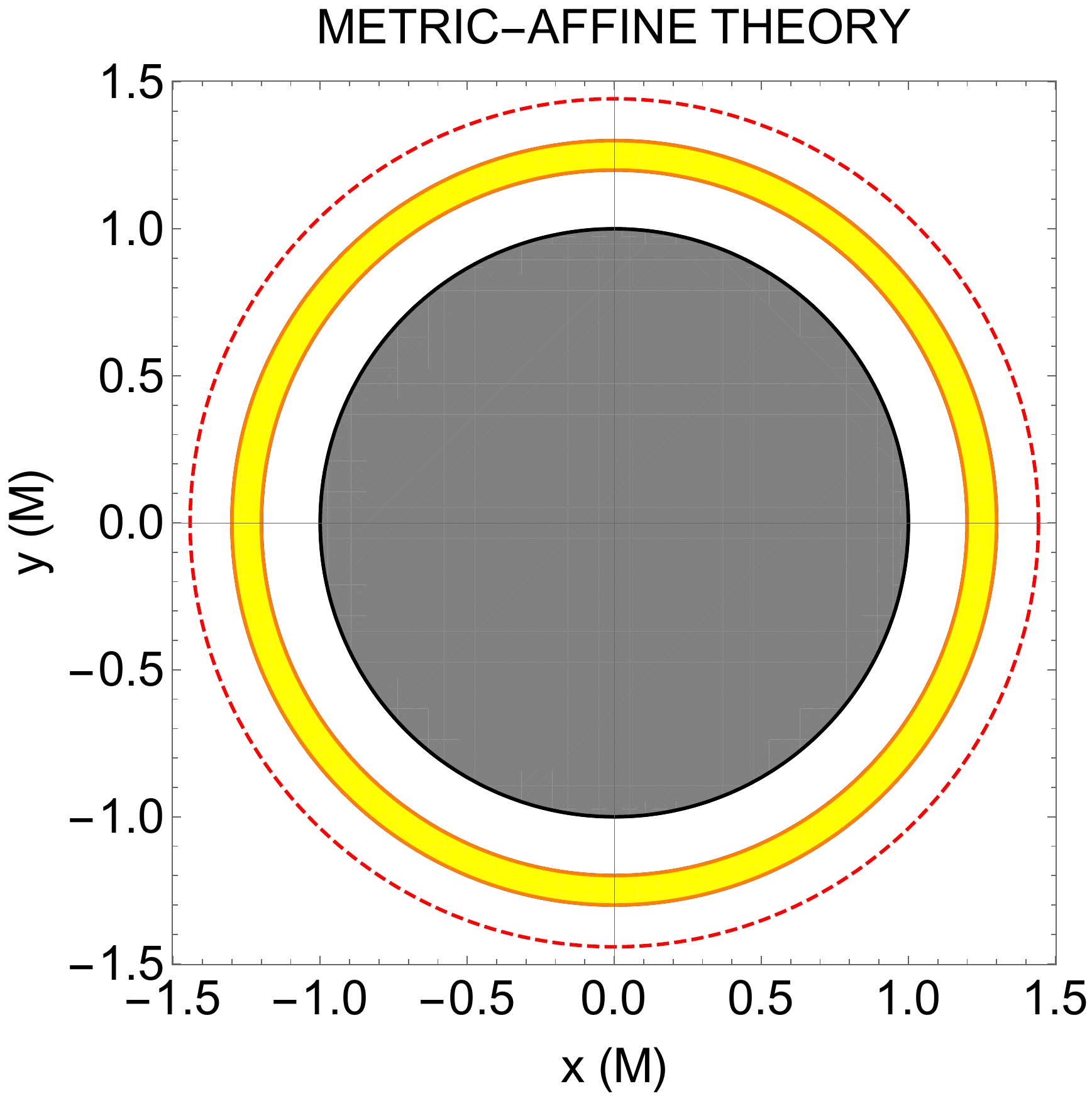}
\hspace{0.55cm}
\includegraphics[scale=0.21]{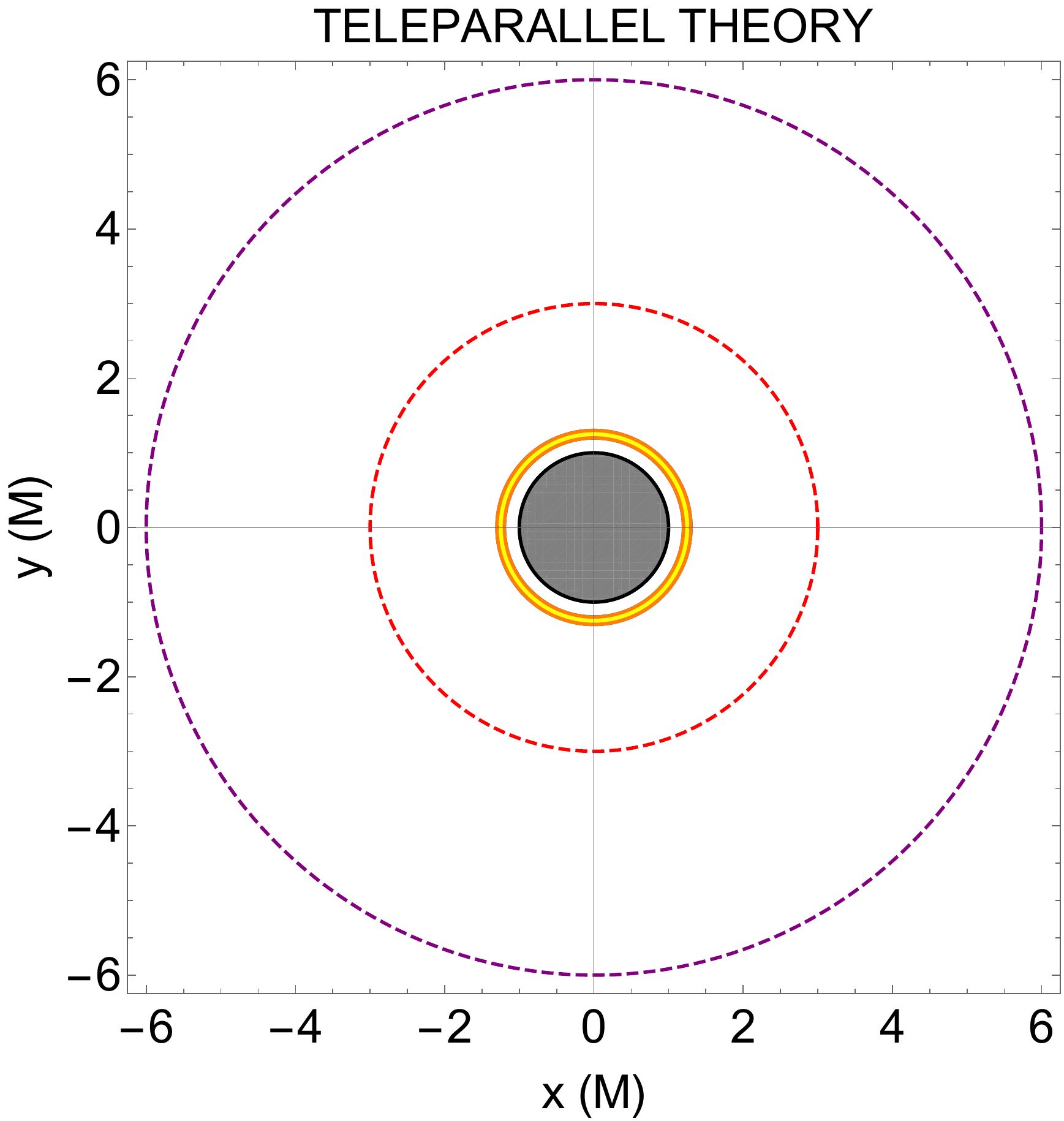}}
\vspace{0.2cm}
%2) Lightcurves
\hspace{-0.3cm}\hbox{
\includegraphics[scale=0.16]{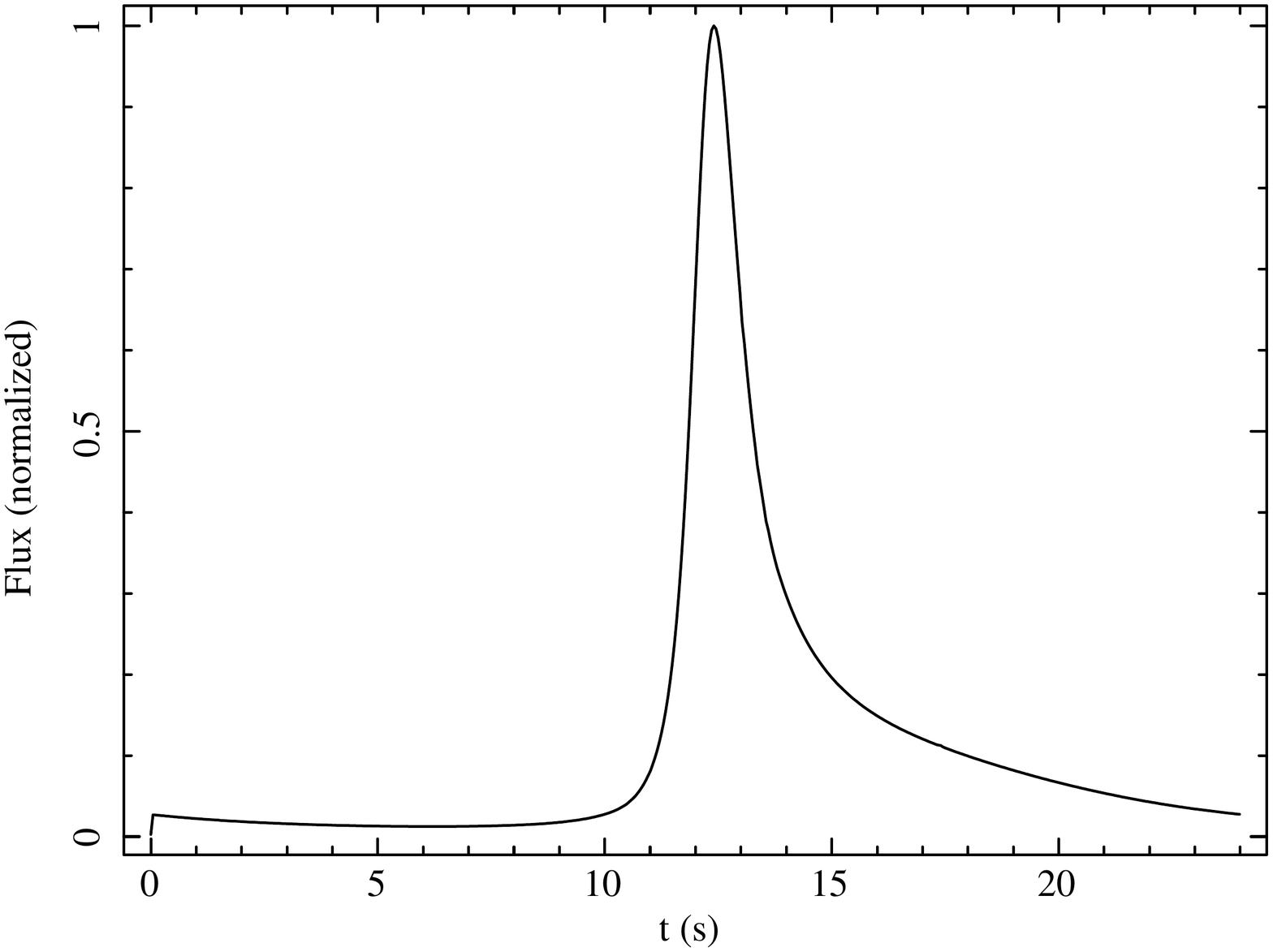}
\hspace{0cm}
\includegraphics[scale=0.16]{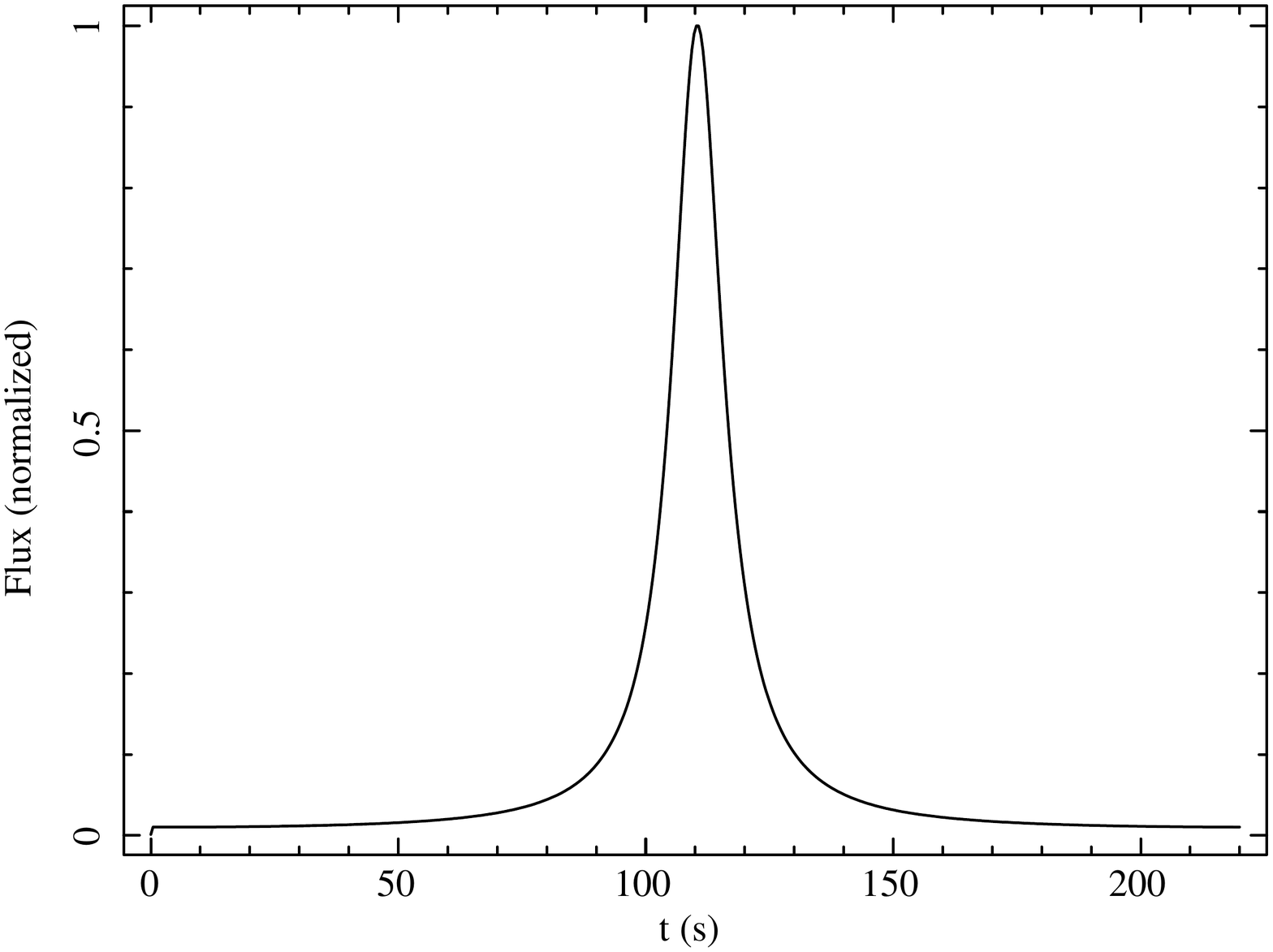}
\hspace{0cm}
\includegraphics[scale=0.16]{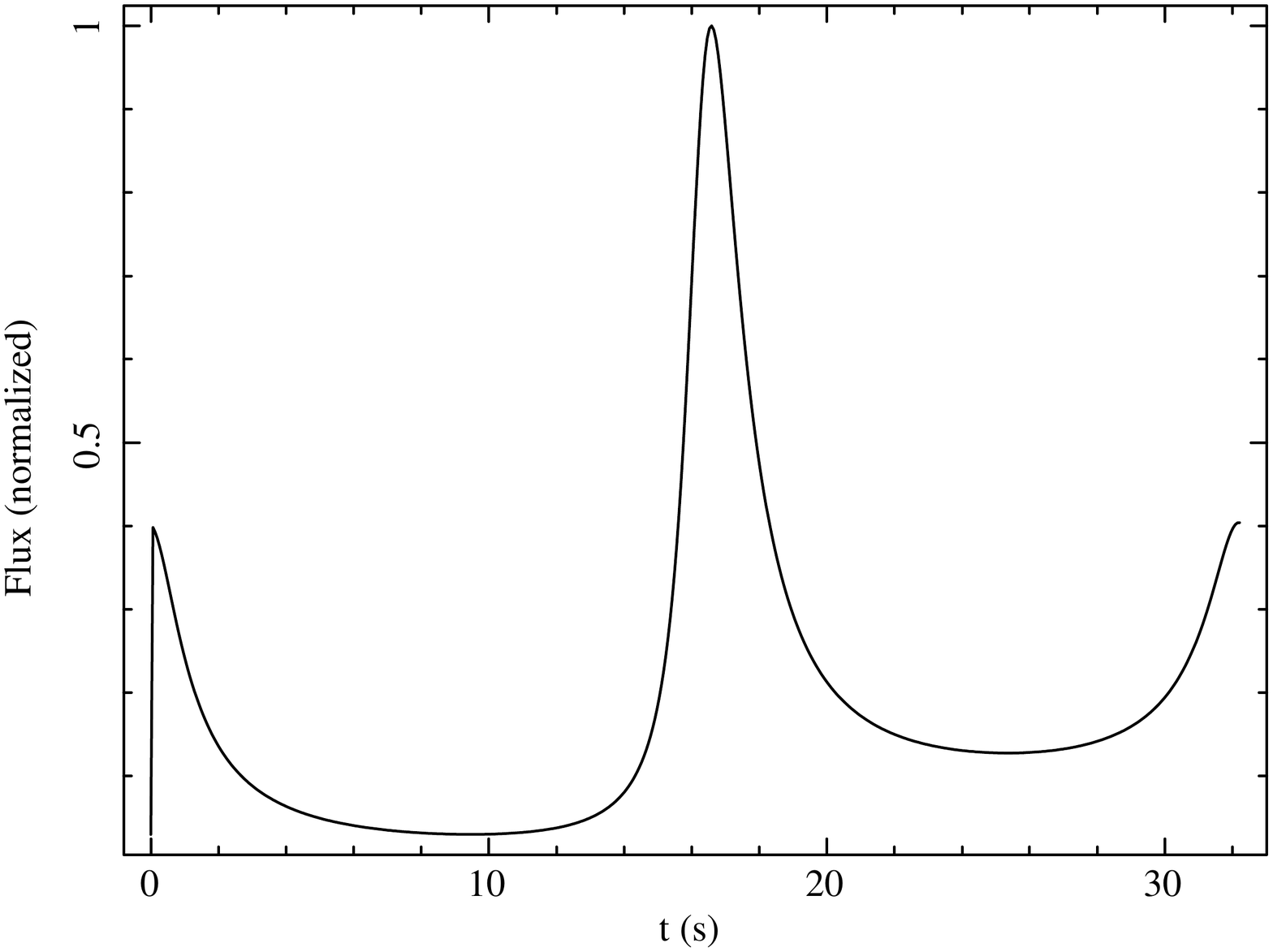}
\hspace{0cm}
\includegraphics[scale=0.16]{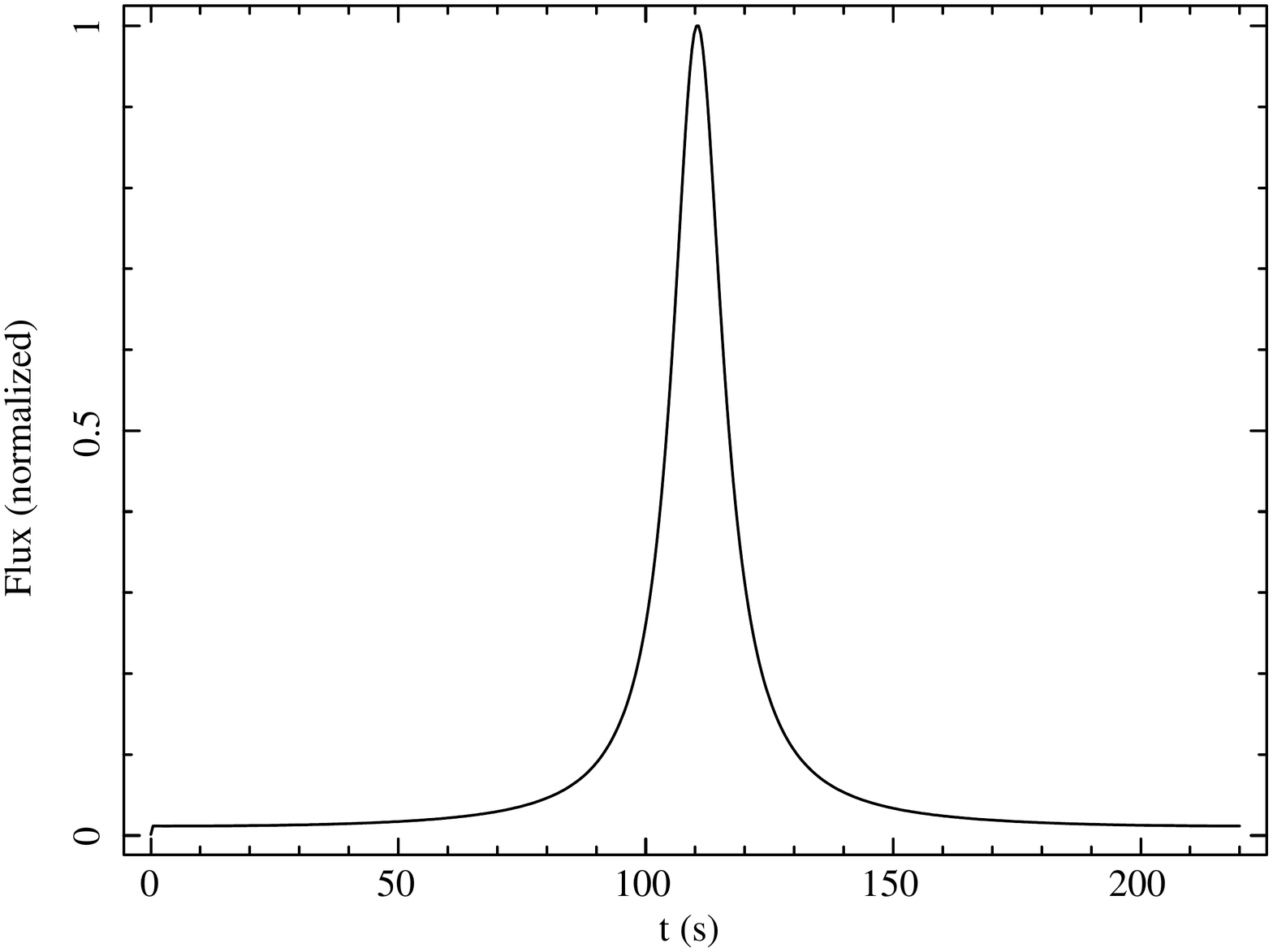}}
\vspace{0.2cm}
%3) Spectra
\hspace{-0.3cm}\hbox{
\includegraphics[scale=0.16]{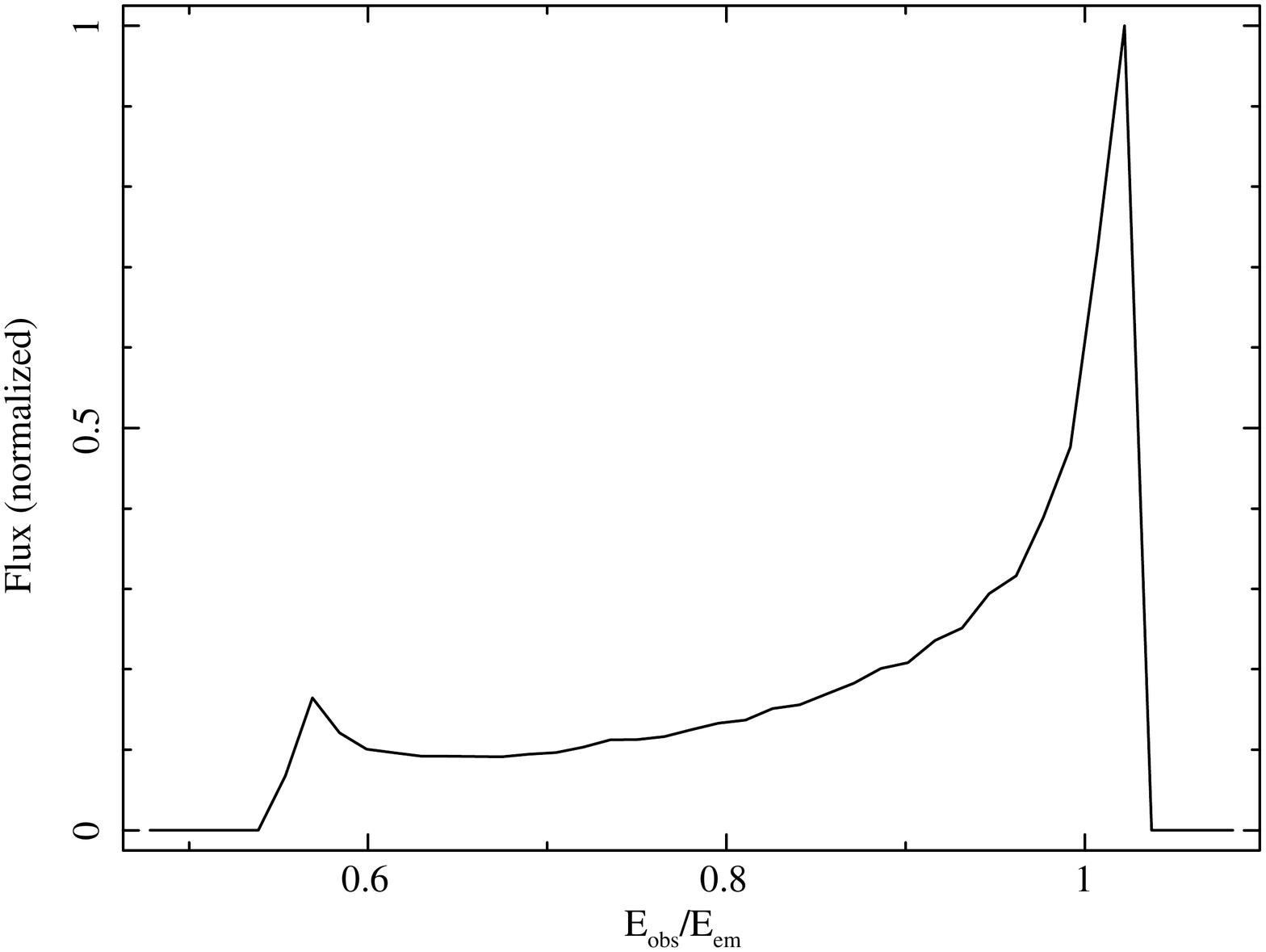}
\hspace{0cm}
\includegraphics[scale=0.16]{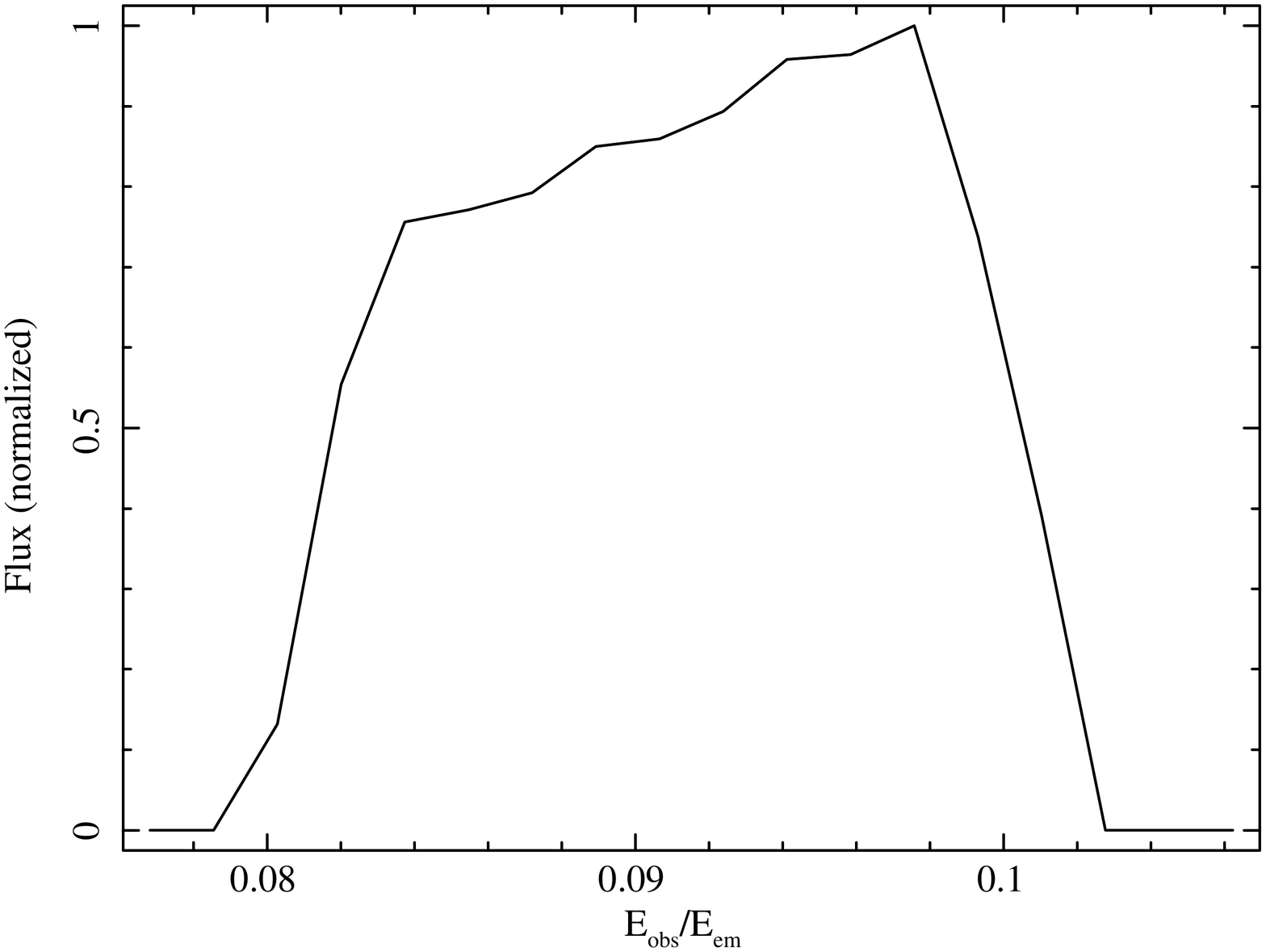}
\hspace{0cm}
\includegraphics[scale=0.16]{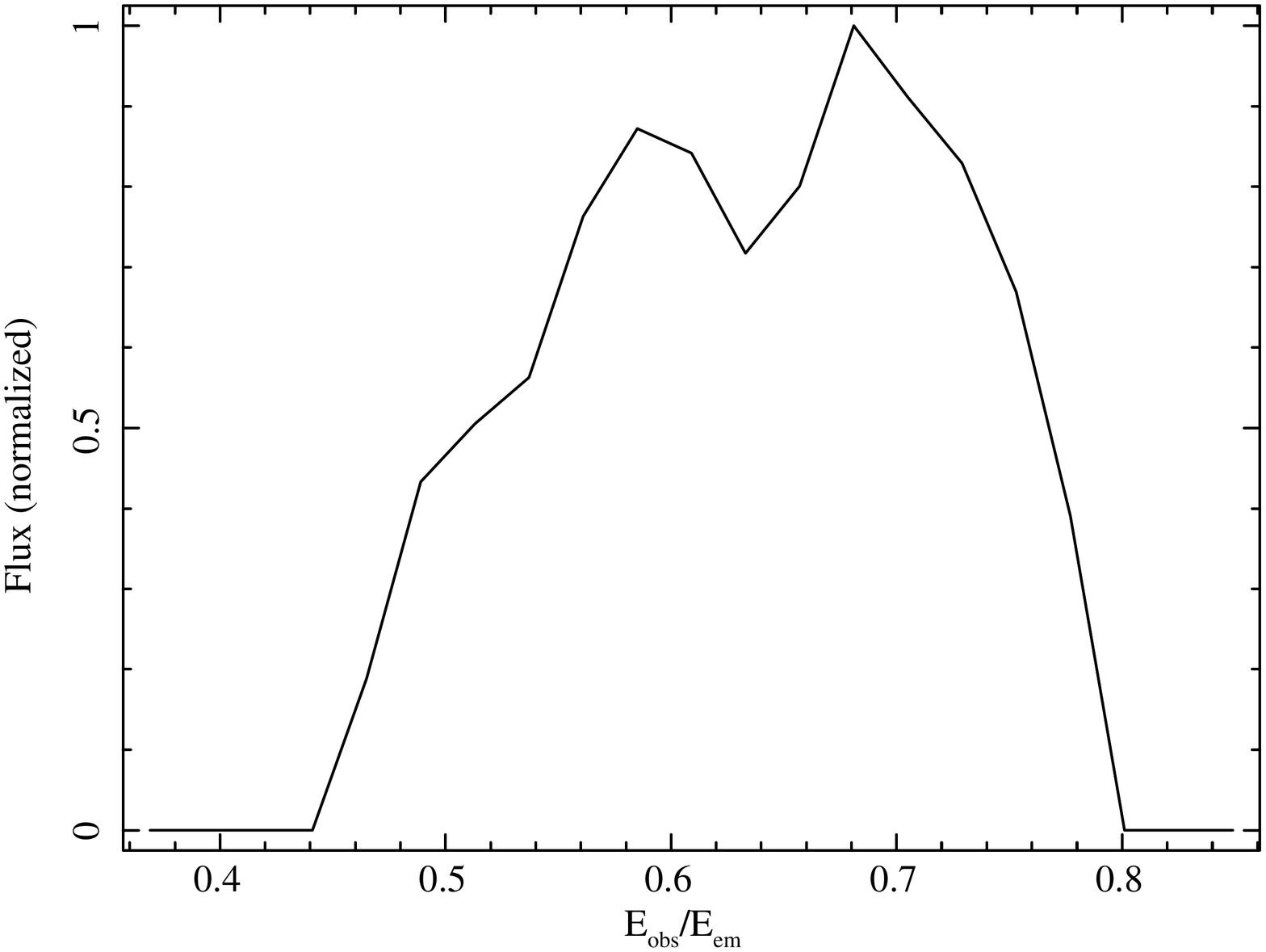}
\hspace{0cm}
\includegraphics[scale=0.16]{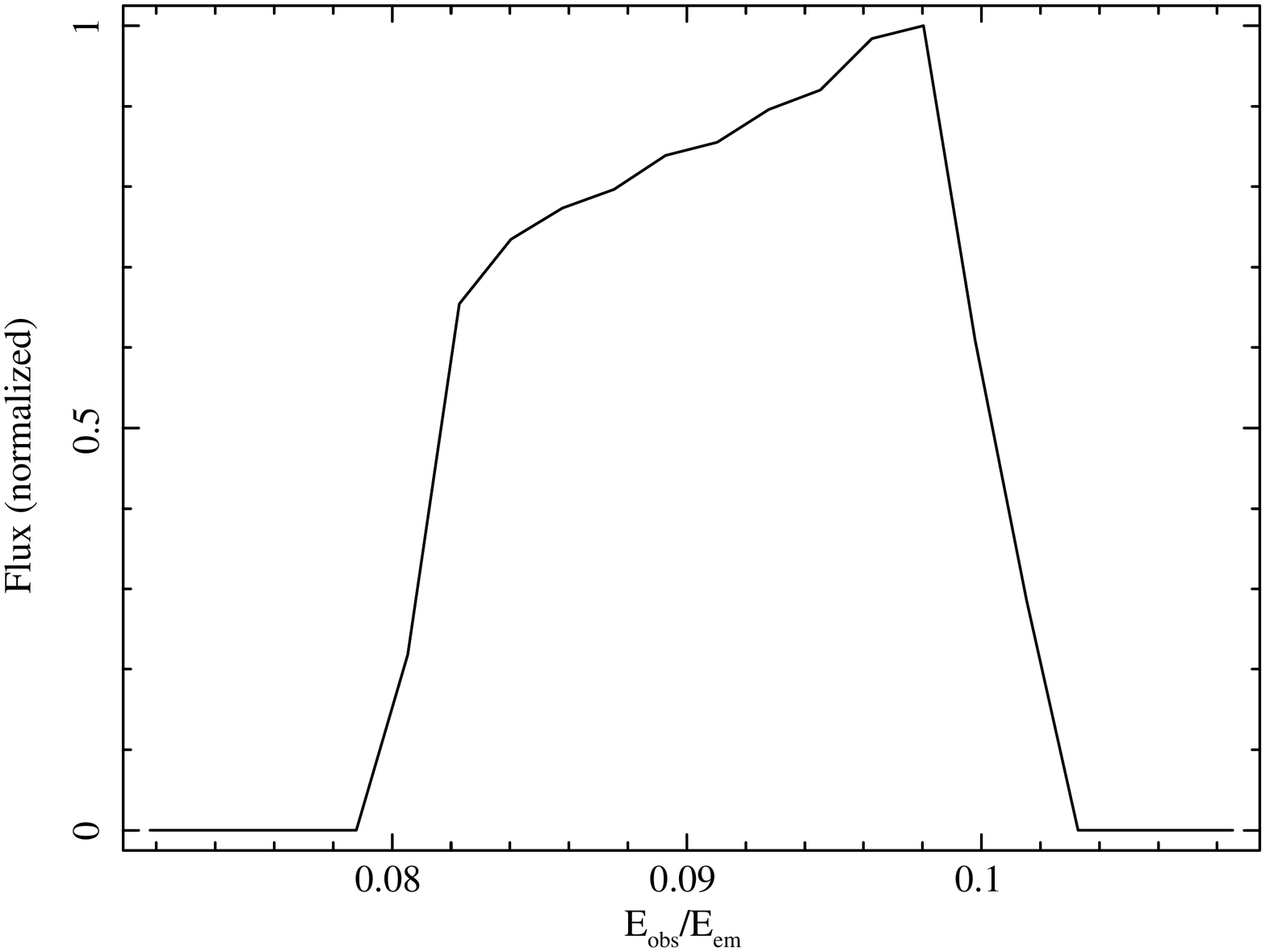}}
\caption{The PR critical hypersurface, extending from $R=1.2M$ to $R=1.3M$ (except in the GR theory that extends from $R=1.2\ b_0$ to $R=1.3\ b_0$) for different WH solutions belonging to the four distinct extended theories of gravity are plotted, considering an observer inclined by an angle $i=80^\circ$.
\emph{First row:} The PR critical hypersurface (yellow surface), the WH throat (dark gray surface), and the photon sphere (dashed red line) and ISCO (violet dashed line) radii are plotted.
\emph{Second row:} The lightcurve of the PR critical hypersurface is plotted.
\emph{Third row:} The spectrum of the PR critical hypersurface is plotted.
}
\label{fig:Fig3}
\end{figure*}
In order to investigate the geometry of the different WH solutions we plot in Fig. \ref{fig:Fig2} their shapes on which we highlight also the behaviour of the redshift function (see Ref. \cite{Morris1988}, for the spatial geometrical visualization of a WH embedded in the 3D Euclidean space). The WH solution obtained in GR theory using the Casimir effect as a stress-energy tensor gives rise to a \emph{micro-WH} of few Planckian size (i.e, $b_0\approx0.37\ell_{\rm P}$). Therefore, it is very challenging to detect it with the actual technologies, because the metric outside the WH throat at distances of order of $M$ is generally flat. However, they are very intriguing because they may hide critical information both on the nature of the spacetime geometry and on gravity at microscopic levels. The other WH solutions clearly show that $\Phi(r),b(r)$ are monotone increasing functions, and the geometry of the WH neck changes in each WH solution. This plot permits to have a fast overview of the WH solutions to quickly learn their proprieties for further theoretical and observational speculations.
\begin{figure}[h!]
\centering
\vbox{
\includegraphics[scale=0.305]{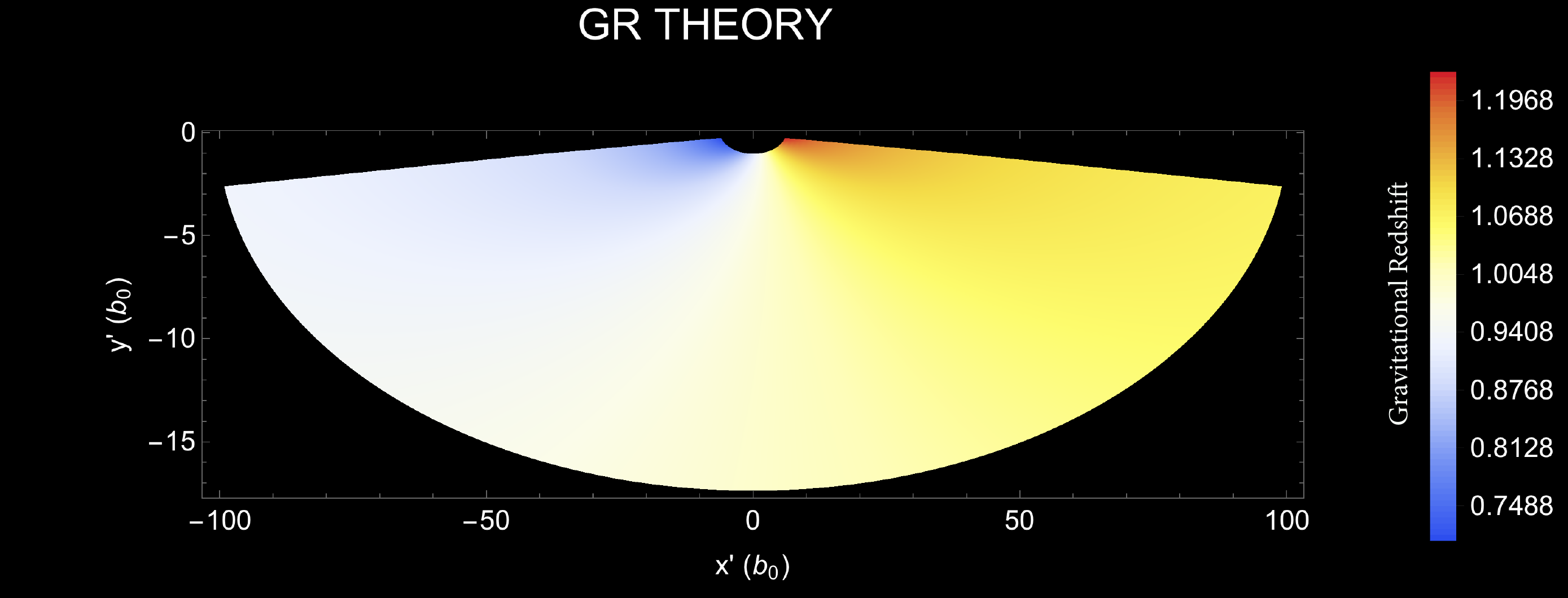}\\
\vspace{-0.1cm}
\includegraphics[scale=0.305]{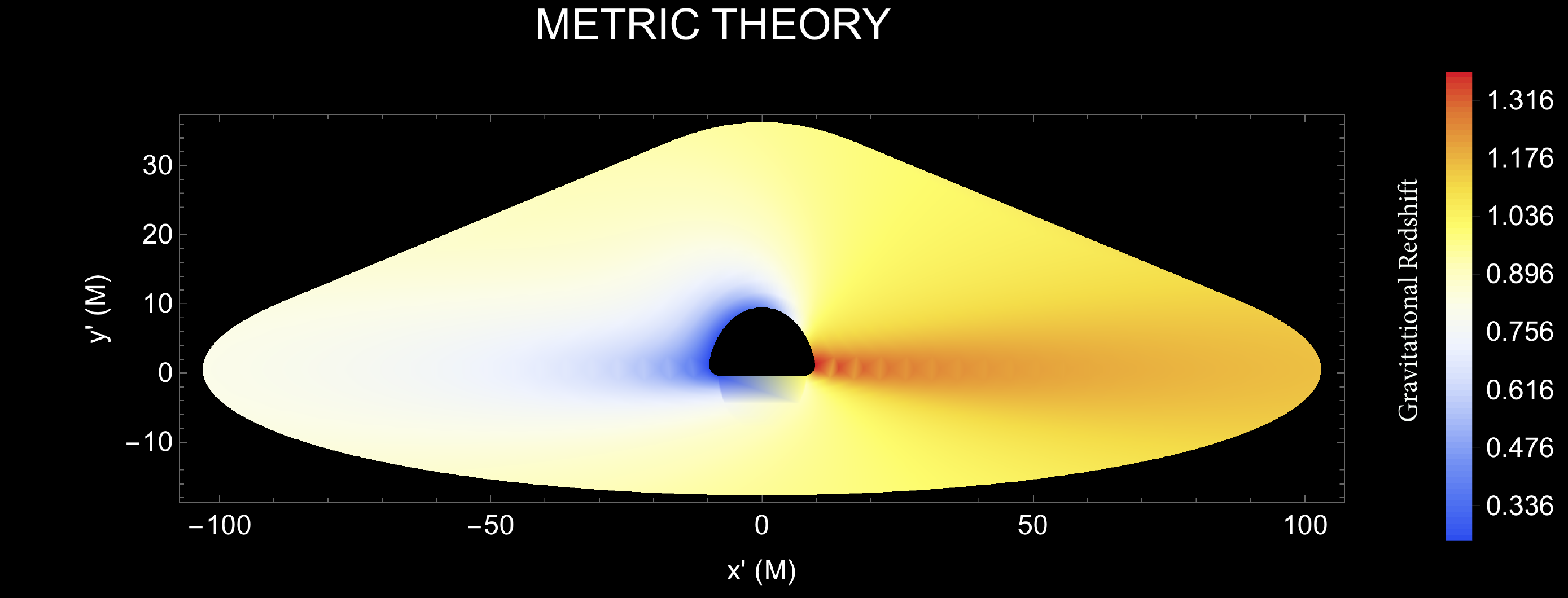}\\
\vspace{-0.1cm}
\includegraphics[scale=0.305]{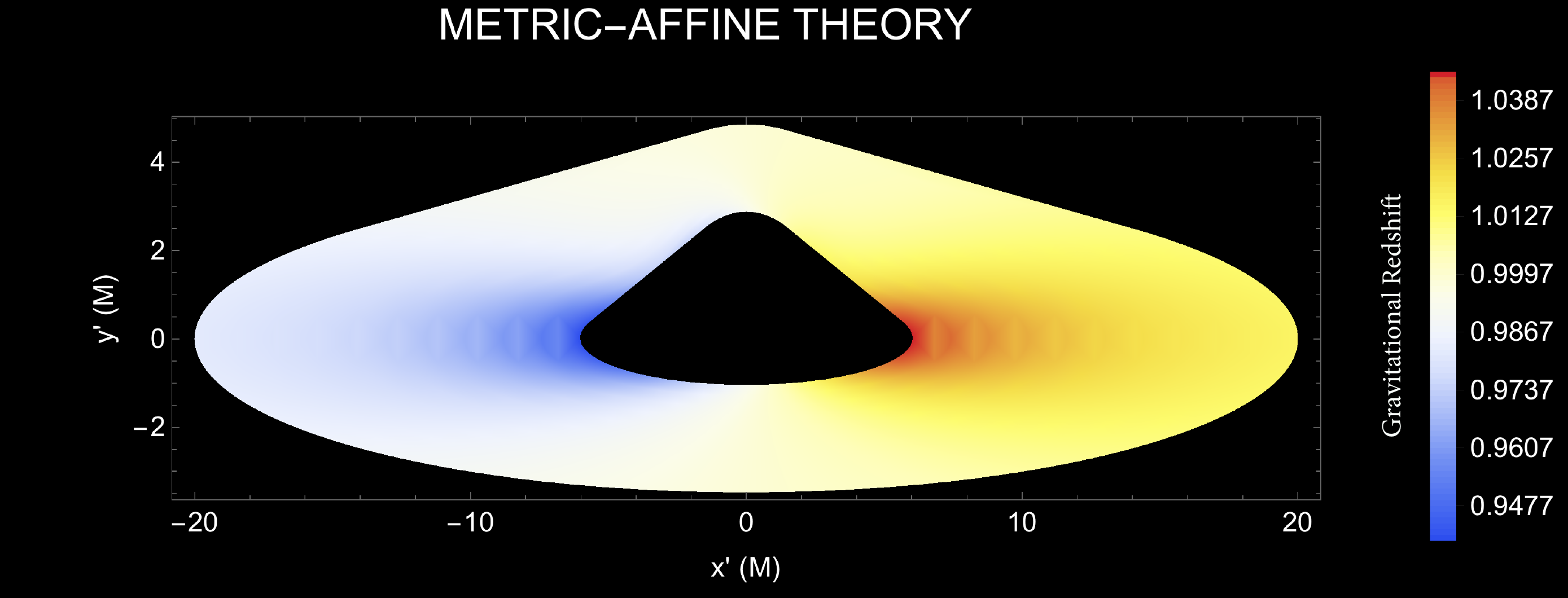}\\
\vspace{-0.1cm}
\includegraphics[scale=0.305]{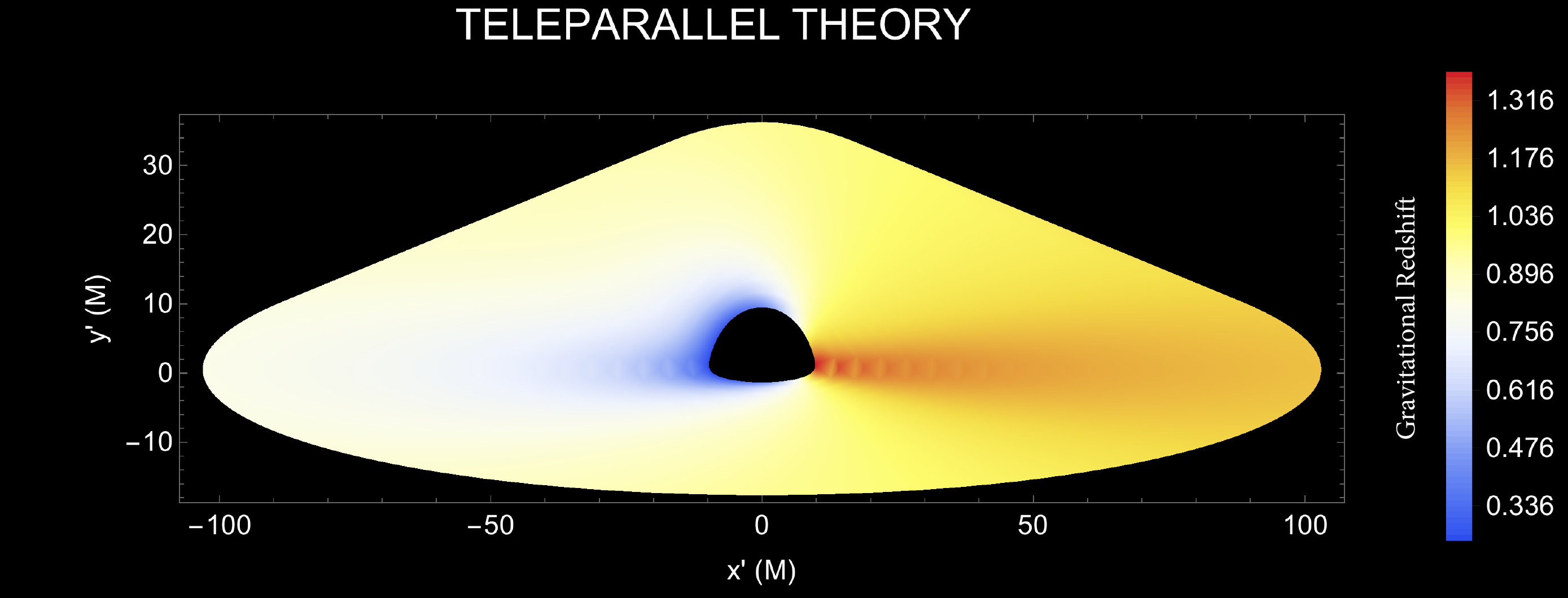}}
\caption{Image of an accretion disk seen by a static, and no-rotating observer at infinity inclined by an angle $i=80^\circ$. The disk extends: from $R_{\rm in}=6M$ to $R_{\rm out}=100M$ for WH solutions in the metric and teleparallel theories; from $R_{\rm in}=6\ b_0$ to $R_{\rm out}=100\ b_0$ for the WH solution in the GR theory; from $R_{\rm in}=6M$ to $R_{\rm out}=20M$ for the WH solution in the metric-affine theory, where the disk is composed by PR critical hypersurfaces in which $\lambda=0.3$ at all radii. The lateral legend bar indicates the level of gravitational redshift in each point of the disk, being different for each WH solution.}
\label{fig:Fig4}
\end{figure}

In Fig. \ref{fig:Fig3} we consider the PR critical hypersurface extending from $R_i=1.2M$ to $R_f=1.3M$, except in GR theory where it spans from $R_i=1.2b_0$ to $R_f=1.3b_0$. In all cases (except in GR theory), the PR critical hypersurface is located under the photon sphere, where possible metric-changes could occur \cite{Defalco2020WH}. We consider that the matter on the PR crtitical hypersurface moves with constant angular velocity. For this reason, we fix $\lambda_0=0.5$ at $R_i=1.2M$ (at $R_i=1.2b_0$ in GR theory) and then we let $\lambda(r)$ vary with $r\in[R_i,R_f]$ according to this law
\begin{equation}
\lambda(r)=\frac{e^{\Phi(R_i)}}{e^{\Phi(r)}}\left(\frac{r}{R_i}\right)^2\lambda_0.  
\end{equation}

In this way we can calculate the intervals over which the PR photon impact parameter $\lambda$ and the luminosity parameter $A/M$ range. We obtain the following results:
\begin{enumerate}
    \item GR theory: $\lambda\in[0.50,0.54]$, $A/b_0\in[0.10,0.12]$; 
    \item Metric theory: $\lambda\in[0.50,0.52]$, $A/M\in[0.01,0.02]$;
    \item Metric-affine theory: $\lambda\in[0.50,0.54]$,  $A/M\in[0.37,0.46]$;
    \item Teleparallel theory:  $\lambda\in[0.50,0.52]$, $A/M\in[0.01,0.02]$.
\end{enumerate}
All WH solutions admit physically reasonable luminosity parameter ranges, except maybe for the WH solution in metric-affine theory which is relatively high.

The lightcurves of the PR critical hypersurfaces are plotted for just one period, time that the test particle takes to complete one loop from the starting position. In these plots, it is possible to examine the dynamical evolution of the emitted fluxes. We note that all plots show almost the same trend, similar to a peaked Gaussian at half period, corresponding to the material on the approaching side toward the observer which is greatly enhanced by the gravitational field. In all cases (except in GR theory), the PR critical hypersurface is located under the photon sphere radius, therefore only the part of the PR critical hypersurface which is located inside the cone of avoidance reaches the observer at infinity. The WH solution in the metric-affine theory shows some visible differences at the begin and end of its periodic motion, while the WH solutions in metric and teleparallel cases exhibit only very tiny differences, which are almost indistinguishable. The explanation of such behaviours relies on the evidence that the redshift functions $\Phi(r)$ of these two theories are equal and dominate over their shape functions, albeit they have different functional forms. Instead in the metric-affine case, it is worth to note that the shape function is always dominated in the blueshifted region, but it contributes at the begin and end of the period by producing small enhancements in the flux.

Another fundamental enquiry relies on the spectral analysis, which investigates how the flux changes in terms of the energy. We note that the spectral profiles of the WH solutions in metric and teleparallel theories are still very similar, and are also defined in the same energy interval, i.e. $E_{\rm obs}/E_{\rm em}\in[0.010,0.104]$. The explanation of such analogy resumes the same argument exposed in the lightcurve case. In the other cases alternately, the profiles are well distinguishable and therefore it is very easy to recognise them. From such example we learn that for some WH solutions it is difficult to reconstruct $\Phi(r)$ and $b(r)$ functions only looking at their PR critical hypersurfaces. Other alternative ways to better disentangle two (or more) similar WH solutions belonging to different or same extended theories of gravity could be: (i) looking for other emitting surfaces or (ii) searching for other strategies which can be combined with that developed here to crosscheck the results and extract new information.

Finally, in Fig. \ref{fig:Fig4} we consider an accretion disk forming around a WH, where we numerically simulate its appearance on the screen of a static and no-rotating observer at infinity. For the WH solution in GR theory we consider a small accretion disk extending from $R_{\rm in}=6\ b_0$ to $R_{\rm out}=100\ b_0$, otherwise if it is too far from the WH throat the results become trivial, because the disk would lie in a quasi flat spacetime. For the WH solutions in metric and teleparallel theories, a real disk extending from $R_{\rm in}=6M$ to $R_{\rm out}=100M$ is considered. Instead, for the WH solution in metric-affine theory it is not possible to have an accretion disk supported uniquely by the background gravitational field, because stable circular orbits do not exist (see Sec. \ref{sec:AFFINE-METRIC}). However, we consider a disk made by PR critical hypersurfaces, extending from $R_{\rm in}=6M$ to $R_{\rm out}=20M$, and endowed with PR photon impact parameter $\lambda=0.3$ in all points of the disk.
This case is not fictitious at all, because it realistically reproduces the physics occurring in the inner regions of an astrophysical disk, which are radiation-pressure dominated (see Refs. \cite{Shakura1973,Lancova2017}, for further details). With our parameter choices, we have that the luminosity parameter ranges in $A/M\in[0.003,0.0067]$. This result gives rise to a counterintuitive issue, because we expect that the more far we move from an emitting source (located close to the WH throat), the stronger the luminosity should be. This controversy can be better understood if we think that the geometrical proprieties of the background spacetime are different from those of classical GR, where the PR effect is normally framed. In the other cases, we consider that the matter moves inside the disk with Keplerian angular velocity (see Table \ref{tab:Table2}, for a summary of the equations employed for obtaining the WH image). In all cases, we plot next to each image a legend bar showing the gravitational redshift in each point of the accretion disk. This permits to understand not only the WH geometrical structure, but also how the matter interacts and behaves with the gravity in each WH spacetime. We immediately note that also in this case, the WH solutions in metric and teleparallel theories are still similar, being hard to spot the differences. Therefore, although the ray-tracing and imaging procedures are very powerful and successful diagnostic tools in several cases, sometimes they should be complemented by alternative strategies in order to achieve more solid and robust results. 

\renewcommand{\arraystretch}{1.8}
\begin{table*}[t!]
\begin{center}
\caption{\label{tab:Table2} Summary of the accretion disk information around the WH solutions. Critical impact parameter, Keplerian velocity, periastron, and gravitational redshift are calculated respectively through Eqs. (\ref{eq:crim}), (\ref{eq:OmK}), (\ref{eq:p}), and (\ref{eq:reddisk}).\\
}	
\normalsize
\vspace{0.3cm}
\begin{tabular}{|@{} c @{}|@{} c @{}|@{} c @{}|@{} c @{}|@{} c @{}|} 
\ChangeRT{1pt}
\ \ {\bf Theory}\ \ &\ \ $\boldsymbol{b_c}$ \ \ & $\boldsymbol{p}$ & $\boldsymbol{\Omega_K}$ & $\boldsymbol{(1+z)^{-1}}$\\
\ChangeRT{1pt}
%GR
GR & $0.66\ b_0$ & NOT EXIST &\hspace{0.5cm} $
4 \sqrt{6} \sqrt{\frac{b_0}{R^2 (3 R+b_0)}} \left(\frac{R}{4 R+b_0}\right)^{3/2}$\hspace{0.5cm} &\hspace{0.5cm} $\frac{8 \left(\frac{R}{4 R+b_0}\right)^{3/2} \sqrt{1-\frac{3 b_0}{2 (3 R+b_0)}}}{1+b_{\rm ph}\Omega_K\frac{\sin i\sin\varphi}{\sin\psi}}$\hspace{0.5cm} \\
\hline
%METRICO
Metric &\hspace{0.1cm} $8.15\ M$ \hspace{0.1cm} &\hspace{0.1cm} $-\frac{3M}{W(-3M/b_{\rm ph})}$\footnote{We remind that $W$ is the Lambert (or \texttt{ProductLog}) function \cite{FUKUSHIMA2013}.} \hspace{0.1cm} &\hspace{0.5cm} $\sqrt{3} e^{-3M/r} \sqrt{\frac{M}{r^3}}$\hspace{0.5cm} &\hspace{0.5cm} $\frac{e^{-3M/r} \sqrt{1-3M/r}}{1+b_{\rm ph}\Omega_K\frac{\sin i\sin\varphi}{\sin\psi}}$\hspace{0.5cm} \\
\hline
%METRICO-AFFINE
\ \ Affine-Metric\ \ &  $2.01M$ & $-\frac{\sqrt[3]{3}M}{\sqrt[3]{W\left[-3\left(\frac{M}{b_{\rm ph}}\right)^3\right]}}$ &
$\sqrt{3} e^{-\frac{M^3}{r^3}} \sqrt{\frac{M^5}{r^5}}$\footnote{It is important to note that in such WH spacetime there not exist stable circular orbits, see Sec. \ref{sec:AFFINE-METRIC}. Therefore, the Keplerian angular velocity and the gravitational redshift (for a disk) are not globally defined. However, we are able to derive their analytical expressions, as those reported in the table, but they can be interpreted only as expressions valid locally.} & $\frac{e^{-\frac{M^3}{r^3}} \sqrt{1-\frac{3M^3}{r^3}}}{1+b_{\rm ph}\Omega_K\frac{\sin i\sin\varphi}{\sin\psi}}$ \\
\hline
%TELEPARALLELO
Teleparallel & $8.15\ M$ &$-\frac{3M}{W(-3M/b_{\rm ph})}$ & $\sqrt{3} e^{-3M/r} \sqrt{\frac{M}{r^3}}$ & $\frac{e^{-3M/r} \sqrt{1-3M/r}}{1+b_{\rm ph}\Omega_K\frac{\sin i\sin\varphi}{\sin\psi}}$ \\
\ChangeRT{1pt}
\end{tabular}
\end{center}
\end{table*}

\section{Conclusions}
\label{sec:end}
This paper configures as a follow up analysis to our previous work \cite{Defalco2020WH}. This paper configures as a follow up analysis to our former work \cite{Defalco2020WH}. Indeed, in our previous publication we employed the flux emitted by the PR critical hypersurfaces, modeled in the BH Schwarzschild metric, through the ray-tracing equations in GR as a probe to detect possible metric changes occurring in the transition surface layer, located between the BH event horizon and the photon sphere. Since this scenario entails the possible existence of a WH, we are naturally led to look for a procedure to distinguish among the different WH solutions which are the most adapt to fit and interpret the observational data.

To solve this issue, we have proposed a strategy for testing static and spherically symmetric WH solutions described by the general Morris-Thorne-like metric (\ref{eq:MTmetric}) against the observational data. This method is based on the following three fundamental steps: 
\begin{enumerate}
\item development of general ray-tracing equations in metric (\ref{eq:MTmetric}), which are: light bending (\ref{eq:LB}), travel time delay (\ref{eq:TD}), and solid angle (\ref{eq:SA}), see Sec. \ref{sec:RTE};
\item employing the general relativistic PR effect model framed in the general metric (\ref{eq:MTmetric}) \cite{Defalco2020WH} to determine the existence of critical hypersurfaces (\ref{eq:CH}) in strong field regime regions, see Sec. \ref{sec:PRE};
\item calculation of the flux (\ref{eq:flux}) emitted by the PR critical hypersurface (or also by an accretion disk in the equatorial plane around a WH and located only in one universe) toward a distant observer. This model is so general and flexible that it is capable to perfectly model all different spherically symmetric and static WH solutions, without using any approximation. This permits to fit all observational data presenting metric-changes, and to consequently determine the profile of the unknown functions $\Phi(r),b(r)$ relative to the region where the emitting surface is placed. The observational data can thus impose tight constraints not only on the WH solutions, but also on the theories of gravity. 
\end{enumerate}

We would like to underline that this work and the previous publication \cite{Defalco2020WH} are conceptually distinct. Moreover, this paper configures as a logical and direct consequence of the previous article. The fundamental differences between the two papers can be summarized as follows: 
\begin{enumerate}
\item \emph{practical intents}, because in the first paper we looked for observational WH existence, while in this manuscript we have constrained the WH solutions within general gravity frameworks through the fit of the data; 
\item \emph{geometrical approach}, because in the first work we only employed the BH Schwarzschild metric, while in this manuscript we have dealt with generic static and spherically symmetric WH metrics; 
\item \emph{philosophy and methodology of investigation}, because in the former publication we have only considered a single  specific metric and we based our conclusions on the dihcotomic answer \qm{yes} or \qm{no} to the question of the WH existence; on the other hand, in this paper we have handled a two-parameters class of general metrics and we have ontologically dealt with several WH solutions, which can be reduced and constrained only through the data, producing as a final result a \qm{procedure}, rather than a \qm{specific answer}.
\end{enumerate}

The fundamental element of this work is represented by metric (\ref{eq:MTmetric}), which has a general character and valid both in GR and in extended/alternative theories of gravity. The ray-tracing technique combined with the imaging procedure permit to infer fundamental information on the WH solutions through the fit of the observational data. In some cases, we are not able to identify just one single WH solution, but we can reconstruct a class of possible WH solutions. To properly determine the right WH solution,
it is necessary to exploit alternative procedures for extracting the missing information, see Sec. \ref{sec:discuss}.

This procedure allows also to provide new tests of gravity in strong field regimes, understanding thus whether GR or extended/alternative theories of gravity are needed. Another advantage of this work is represented by the general set of ray-tracing equations, which can be employed to calculate and model the flux from any x-ray source, regardless of its geometry. The same argument holds also for any configuration of the observer's geometrical screen structure and dynamical setup. 

To show the potentiality of our approach, we have applied the general ray-tracing equations, see Sec. \ref{sec:geometry}, to four different WH solutions belonging to distinct extended theories of gravity, which are: GR (see Sec. \ref{sec:GR}), metric (see Sec. \ref{sec:METRIC2}), metric affine (see Sec. \ref{sec:AFFINE-METRIC}), and teleparallel (see Sec. \ref{sec:TELEPARALLEL}). For each WH solution we have analysed the geometry close to the WH neck (see Fig. \ref{fig:Fig2}), the lightcurve and spectrum of the PR critical hypersurface around a WH toward a distant observer (see Fig. \ref{fig:Fig3}), and the image of an accretion disk seen by an observer at infinity (see Fig. \ref{fig:Fig4}).  We have conducted an accurate study on these WH solutions, showing how to extract important information on the WH geometrical proprieties (see Tables \ref{tab:Table1} and \ref{tab:Table2}, for more details). In particular the calculus of the photon sphere and ISCO radii are important a-priori checks for having a quick overview of a WH spacetime's geodesic structure. 

A limiting case of this model is represented by the fact that all the ray-tracing equations are \emph{elliptic integrals}, which do not admit a solution in terms of elementary functions \cite{Defalco2016}. However, the calculation of such integrals strongly depends also on the functional form of $\Phi(r),b(r)$, but generally, they give rise to complicate functions, which are very time-consuming. However, following the same ideas adopted in the Schwarzschild case, it would be helpful to develop a set of high-accurate approximate equations (see Refs. \cite{Defalco2016,Semerak2015,Laplaca2019,Poutanen2019}, for more details), which drastically reduce the  computational times. Our ray-tracing approach could be further complemented by the model-independent framework developed by Rezzolla and collaborators \cite{Rezzolla2014}, where through a finite set of coefficients it would be possible to measure possible metric-deviations from GR for all static and spherically symmetric geometries mimicking that of a BH. 

As future perspectives, we aim at extending this treatment also to axially symmetric spacetimes in the equatorial plane and then in the 3D space.

\section*{Acknowledgements}
V.D.F. thanks Gruppo Nazionale di Fisica Matematica of Istituto Nazionale di Alta Matematica (INDAM) for support. S.C. acknowledges the support of Istituto Nazionale di Fisica Nucleare (INFN) sez. di Napoli, Iniziative Specifiche MOONLIGHT2 and QGSKY. M.D.L. acknowledges INFN sez. di Napoli, Iniziative Specifiche QGSKY and TEONGRAV.

\bibliography{references}
\end{document}